\documentclass[a4paper,10pt]{article}
\usepackage{amsmath}
\usepackage{graphicx}
\usepackage{caption}
\usepackage{subcaption}
\usepackage{float}
\usepackage{amssymb}
\usepackage{setspace}
\usepackage[T1]{fontenc}
\usepackage{color}
\usepackage{soulutf8}
\usepackage[utf8]{inputenc} 

\makeatletter
\let\@fnsymbol\@arabic
\makeatother

\author{E.G.Rens{\thanks{Life Sciences, CWI, Science Park 123, 1098XG Amsterdam, The Netherlands}\space$^{,}$\thanks{Mathematical Institute, Leiden University, Niels Bohrweg 1, 2333CA Leiden, The Netherlands}} \space and  
  R.M.H.Merks\footnotemark[1]\space$^{,}$\footnotemark[2]}
\title{title: Cell contractility facilitates alignment of cells and tissues to static uniaxial stretch \\ running title: contractility facilitates cell alignment}
\date{}

\newcommand{\uli}[1]{\smash{\underline{#1}}}

\begin{document}

\maketitle

\newpage

\begin{abstract}
During animal development and homeostasis, the structure of tissues, including muscles, blood vessels and connective tissues adapts to mechanical strains in the extracellular matrix (ECM). These strains originate from the differential growth of tissues or forces due to muscle contraction or gravity. Here we show using a computational model that by amplifying local strain cues, active cell contractility can facilitate and accelerate the reorientation of single cells to static strains. At the collective cell level, the model simulations show that active cell contractility can facilitate the formation of strings along the orientation of stretch. The computational model is based on a hybrid cellular Potts and finite-element simulation framework describing a mechanical cell-substrate feedback, where: 1) cells apply forces on the ECM, such that 2) local strains are generated in the ECM, and 3) cells preferentially extend protrusions along the strain orientation. In accordance with experimental observations, simulated cells align and form string-like structures parallel to static uniaxial stretch. Our model simulations predict that the magnitude of the uniaxial stretch and the strength of the contractile forces regulate a gradual transition between string-like patterns and vascular network-like patterns. Our simulations also suggest that at high population densities, less cell cohesion promotes string formation. \end{abstract}

\section*{Introduction}
During embryonic development, a single fertilized egg cell grows into a complex functional organism \cite{Friedl2009}. Even after years of studying morphogenesis, the organization of cells into tissues, organs and organisms, it still remains a puzzle how cells migrate and form the right pattern in the right part of the body at the right moment \cite{Reig2014}. Apart from chemical signals \cite{Rogers2011}, mechanical signals play an equally important role in morphogenesis \cite{Siedlik2015,Shawky2015}.
Static strains originating from differential growth of tissues are instrumental for the organization of cells in tissues \textit{in vivo}. For example, in quail heart, the endocardium generates strains to which cardiomyocyte microtubules orient \cite{Garita2011}. Wing hinge contractions in \textit{Drosophila} cause anisotropic tension in the wing-blade epithelium, to which the cells align \cite{Aigouy2010}. Using a multiscale computational modeling approach, here we unravel how static strains, {\em e.g.}, resulting from the differential growth of tissues, may drive the organization of cells and tissues. 

\textit{In vitro} and \textit{in silico} experiments have helped to unravel the cellular mechanisms underlying the adaptation of tissues to strain. Myocytes \cite{Collinsworth:2000cq}, mesenchymal stem cells \cite{Liu2014}, muscle cells, and endothelial cells \cite{VanderSchaft2011} orient in parallel to uniaxial static stretch. Furthermore, fibroblasts organize into string-like structures in parallel to the stretch orientation \cite{Eastwood:1998cmc}, whereas endothelial cells form monolayers of cells oriented in parallel to the stretch \cite{VanderSchaft2011}.  

Active cell traction forces play a crucial role in the alignment of cells to static uniaxial stretch. Using contact guidance, cells can adjust their orientation to the fibers which align with strain \cite{Vader2009,Chaubaroux2015}. Then, by pulling on the matrix, cells can further align the fibers \cite{Klebe1990}. Such mechanical cell-fiber feedback can coordinate cell alignment \cite{Takakuda:1996gl,Reinhardt:2013jbe,Reinhardt2014} and string formation \cite{Byrne2015} along strain. However, \textit{in vitro} observations suggest that cell alignment to uniaxial stretch may not necessarily be driven by fiber alignment. Mesenchymal stem cells align along the orientation of strain on a \textit{nonfibrous} matrix \cite{Liu2014}. In stretched collagen matrices, fibroblasts were found to align along strain in the absence of fiber alignment \cite{Eastwood:1998cmc,Tondon:2014po}. Other authors observed that collagen fibers aligned only after the cells had aligned \cite{Lee2008,Pang2011}. Moreover, fibroblasts can orient along the uniaxial stretch even if fibronectin fibers were aligned perpendicular to the stretch \cite{Mudera:2000cmc}. Altogether, these results suggest that cells can orient to stretch independently of the fiber orientation. 

Mathematical modeling is a helpful tool to explore what biophysical mechanisms can explain the alignment of cells to strain. Previous mathematical models \cite{Bischofs2003,De:2007kf} were based on optimization principles. Bischofs and Schwarz \cite{Bischofs2003} proposed that cells minimize the amount of work needed for contracting the matrix. For dipolar cells, the work was minimized if they oriented in parallel with the uniaxial stretch. If the cells were assumed to generate strains in their local environment, cells formed strings, which aligned with an external strain field \cite{Bischofs2003,Bischofs2005,Bischofs2006}. Based on the observation that cells reorganize focal adhesions and stress fibers to maintain constant local stresses, De {\em et al.} \cite{De:2007kf} proposed that cells adapt their contractility and orientation in order to find the minimal local stress in the matrix. They showed that the local stress becomes minimal if a dipolar cell orients in parallel to uniaxial stretch, as in this configuration the cell traction forces counteract the uniaxial stretch.

In this work, we explain cellular alignment to strain based on a mesoscopic, experimentally testable cellular mechanism. To simulate this mechanism, we propose a hybrid computational model in which the Cellular Potts Model (CPM) \cite{Graner1992} is coupled to a finite-element model (FEM) of the matrix. The computational model \cite{Oers:2014pcb} captures the mechanical cross-talk between the extracellular matric (ECM) and the cells as follows: 1) cells apply forces on the ECM \cite{Lemmon:2010ju}; 2) the resulting strains in the ECM are calculated using a Finite Element Method (FEM); and 3) cells extend protrusions oriented along strain \cite{Pang2011}.  

Based on experimental observations of fibroblasts on elastic substrates \cite{Lo:2000cj} and on modeling studies \cite{De:2007kf}, it has been suggested that cellular traction forces may dominate over, or even counteract global strain cues. Paradoxically, our model suggests that contractile cells locally increase the global uniaxial strain which {\em facilitates} cell alignment to static uniaxial stretch. Our model also suggests that by contracting the matrix, cells can form strings in parallel to the orientation of uniaxial stretch. Finally, our simulations show that differences in cell cohesion and population density may determine under what conditions cells form strings, and when they only align on a individual level. 

\section*{Methods}
\begin{figure}[H]
\centering
\includegraphics{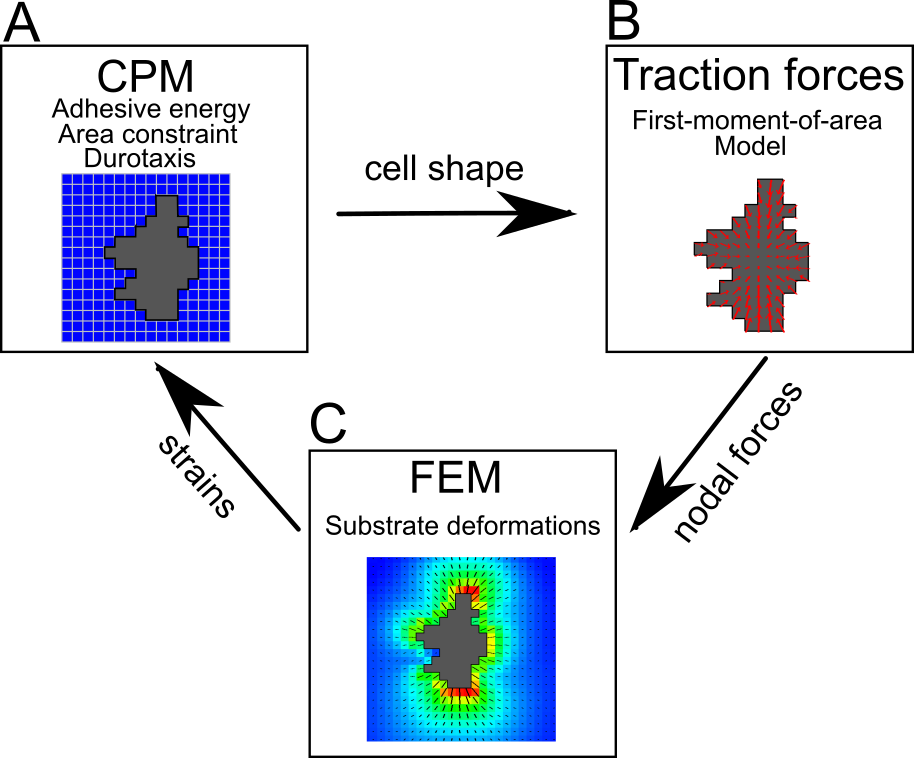}
\caption{Structure of the coupled CPM-FEM model. (A) CPM calculates cell shapes in response to local ECM strains; (B) calculation of cellular traction forces based on cell shapes \cite{Lemmon:2010ju}; (C) substrate strains due to cellular traction forces. }
\label{fig:modeloverview}
\end{figure}

We extended our previous hybrid, cell-based and continuum model \cite{Oers:2014pcb} of mechanical cell-ECM feedback to include the effects of static strain. Figures~\ref{fig:modeloverview}(A-C) give an overview of the model structure. Active cell motility is simulated using the Cellular Potts Model \cite{Graner1992}. The CPM is coupled to a finite-element method that is used to calculate substrate deformations. A time step of the simulation proceeds as follows. Based on the local strains in the matrix and the interactions with adjacent cells, the CPM calculates the cell shapes (Figure~\ref{fig:modeloverview}A). Based on the cell shapes, the traction forces that cells apply on the ECM are determined using the empirically validated first-moment-of-area (FMA) model, as proposed by Lemmon \& Romer \cite{Lemmon:2010ju} (Figure~\ref{fig:modeloverview}B). The FEM calculates the deformation of the substrate resulting from these forces (Figure~\ref{fig:modeloverview}C). Subsequently, the strains in the ECM influence cell movement in the CPM. More precisely, we assume that cells preferentially extend along the orientation of high strain.

\subsection*{Cellular Potts Model}
\label{sec:cpm}
The CPM \cite{Graner1992} describes cells on a regular lattice $\Lambda\subseteq\mathbb{Z}^2$ as a domain of connected lattice sites, $\vec{x}$, of identical spin, or {\em cell identifier}, $\sigma(\vec{x})\in\mathbb{Z}_{\geq 0}$. Sites of spin $\sigma(\vec{x})>0$ identify sections of the substrate that are covered by a biological cell, whereas sites of spin $\sigma(\vec{x})=0$ identify exposed substrate sites. The configuration of cells evolves according to the Hamiltonian,
\begin{equation}
H= \sum_{s\in\mathrm{cells}}\lambda\left(\frac{a(s)-A(s)}{A(s)}\right)^2+\sum_{(\vec{x},\vec{x}\prime)} J(\sigma(\vec{x}),\sigma(\vec{x}\prime))(1-\delta(\sigma(\vec{x}),\sigma(\vec{x}\prime))). \label{eq:H}
\end{equation}
The first term is a surface area constraint, with $a(s)=|\{\vec{x}|\vec{x}\in\Lambda\wedge\sigma(\vec{x})=s\}|$, the number of lattice sites covered by cell $s$,  $A(s)$ a target area and $\lambda$ a Lagrange multiplier. The second term represents the interfacial energies in the system, {\em e.g.}, due to cell adhesion and cortical tensions. Here, $J(\sigma(\vec{x}),\sigma(\vec{x}\prime)$) is the interfacial energy between an adjacent lattice site pair $(\vec{x},\vec{x}\prime)$ and $\delta(\sigma(\vec{x}),\sigma(\vec{x}\prime))$ is the Kronecker delta. The contact energy $J_{cc}$ regulates the degree of cell-cell adhesion, with lower values of $J_{cc}$ corresponding to strong cell-cell adhesion.

To mimic cellular protrusions and retractions of the cells, the cellular Potts model iteratively picks a random lattice site $\vec{x}$ and attempts to copy its spin $\sigma(\vec{x})$ into an adjacent site $\vec{x}\prime$.  The algorithm then calculates the energy change $\Delta H$ associated with the copy attempt based on the Hamiltonian (Eq.~\ref{eq:H}) and any additional energy changes associated with the copy direction \cite{Glazier2007}, in this case $\Delta H_\mathrm{dir}$. With $\Delta H_\mathrm{dir}$ we express the cellular response to matrix strains, as outlined below. The copy is accepted if $\Delta H+\Delta H_\mathrm{dir}\leq0$, or with Boltzmann probability $P(\Delta H+\Delta H_\mathrm{dir})=\exp(-(\Delta H+\Delta H_\mathrm{dir})/T)$ to allow for stochasticity of cell movements.  $T \geq 0$ is a {\em cellular temperature} whose magnitude gives the amount of random cell motility. An additional connectivity constraint  rejects copy attempts that would split cells into disconnected patches.  During one Monte Carlo Step (MCS) $N$  copy attempts are made, with $N=|\Lambda|$, {\em i.e.}, the number of sites in the lattice.

To simulate the response of cells to strains in the substrate, we assumed that local strains promote cellular protrusion and inhibit cellular retractions.  Such a mechanism is motivated by focal adhesions, large integrin complexes that bind the cell to the matrix and maturate on stiffer matrices \cite{Pelham1998}. We assume a strain stiffening material, so that focal adhesions mature on highly strained areas. We thus set 
\begin{equation}
\Delta H_{\mathrm{dir}}= - g(\sigma(\vec{x}),\sigma(\vec{x}\prime)) \lambda_\mathrm{strain}\left(h(E(\epsilon_1))(\vec{v}_1 \cdot \vec{v}_m)^2+h(E(\epsilon_2))(\vec{v}_2\cdot \vec{v}_m)^2\right),
\label{eq:strain}
\end{equation}
where $\lambda_\mathrm{strain}$ is a parameter that describes the sensitivity of cells to strain.  $\vec{v}_m=\widehat{\vec{x}-\vec{x}\prime}$, is the direction of copying, and $\epsilon_1$ and $\epsilon_2$, and $v_1$ and $v_2$  are the eigenvalues and eigenvectors of $\uli{\epsilon}$ that represent the principal strains and strain orientation in the target site $\vec{x}\prime$. We use $g(\sigma(\vec{x}),\sigma(\vec{x}\prime))=1$ if a cell is extending and $g(\sigma(\vec{x}),\sigma(\vec{x}\prime))=-1$ if a cell is retracting, to impose that strain stiffening of the matrix promotes extensions and inhibits retractions. At cell-cell interfaces we assume that the forces due to strain ($\Delta H_{\mathrm{dir}}$) on the extending cell and the retracting cell are balanced, {\em i.e.}, $g(\sigma(\vec{x},\vec{x}\prime))=0$ if $\sigma(\vec{x}) \neq \sigma(\vec{x}\prime)$ and $\sigma(\vec{x})>0 \wedge \sigma(\vec{x}\prime)>0$. We thus assume that neither of the two cells involved in the copy attempt benefits more from occupying a strained lattice site than another cell. The sigmoid function $h(E(\epsilon))= 1/(1+\exp (-\beta(E(\epsilon)-E_\theta)))$ expresses that a minimum stiffness, $E_\theta$, is required for focal adhesion maturation. We assume that cells perceive strain stiffening of the matrix, described by the function $E(\epsilon) = E_0 ( 1 + (\epsilon /  \epsilon_{st} )$, where $\epsilon_{st}$ is a stiffening parameter. Compared to our previous implementation of this model \cite{Oers:2014pcb}, slight adaptations have been made in the Hamiltonian, which are discussed in the Supporting Material. They do not affect the qualitative behavior of the model. The parameter values used in this study are reported in Table S1. We use a discretization of $\Delta x = 2.5 \mu$m. Based on single cell dispersion rates in our model, we previously estimated the time interval $\Delta t$ corresponding to one MCS to be between $\Delta t=0.5$ seconds and $\Delta t =3$ seconds \cite{Oers:2014pcb}.

\subsection*{Finite-element model of compliant substrate}
\label{sec:fem}
A FEM \cite{Davies:2011wr} is used to calculate the strain on the substrate resulting from forces applied to the substrate. The substrate is assumed to be isotropic and linearly elastic. For simplicity, we applied infinitesimal strain theory, assuming that material properties, including local density and stiffness are unchanged by deformations. So, the strain tensor $\epsilon$ is given by
\begin{equation}
\epsilon = \begin{pmatrix} \epsilon_{xx} & \epsilon_{xy} \\ \epsilon_{yx} & \epsilon_{yy} \end{pmatrix} \approx \frac{1}{2} ( \nabla \vec{u} + \nabla \vec{u}^T), 
\end{equation}
where $\vec{u} = (u_x,u_y)$ is the substrate deformation.

The elements of the FEM coincide with the lattice sites of the CPM, {\em i.e.,} the deformation in a lattice site $\vec{u}^e(x,y)$ is approximated by an interpolation of the shape functions $N^e_n(x,y)$, for $n = 1,2,3,4$ corresponding to the four nodes (corners) of lattice site/element $e$:
\begin{equation}
\vec{u}^e(x,y) = \sum_{n=1}^4 N_n(x,y) \vec{u}_n,
\end{equation}
where $\vec{u}_n$ is the substrate deformation at node $n$. We used conventional linear shape functions for four-noded quadrilateral elements \cite{Davies:2011wr}. The FEM is iterated until equilibrium ($K\vec{u}=\vec{f}$, where $K$ denotes the stiffness matrix \cite{Davies:2011wr}), to calculate the deformation $\vec{u}_n$ at each node. All figures in this paper show the strain in equilibrium. The terms $\epsilon_{ij}$ in the strain tensor $\epsilon^e$ of element $e$ are thus given by
\begin{equation}
\epsilon^e_{ij}(x,y) = \frac{1}{2}\sum_{n=1}^4 \frac{\partial N_n(x,y)}{\partial i} \vec{u}_n + \frac{\partial N_n(x,y)}{\partial j} \vec{u}_n.
\end{equation}
In our simulations, the unstretched substrate $\vec{u}=\vec{0}$ is used as a reference configuration for the displacements due to uniaxial stretch and cell contractility. This simplifies our calculations and speeds them up. For details, see Ref.~\cite{Oers:2014pcb}.

\subsection*{Cellular traction forces}
\label{sec:lemmon}
To model the traction forces that cells apply on the substrate we make use of an experimentally-validated, predictive model, called the first-moment-of-area (FMA) model \cite{Lemmon:2010ju}. The FMA model is based on the assumption that the network of actin fibers acts in the cells as a single, cohesive unit.  In the context of our hybrid CPM-FEM model, we implement the FMA model as follows. Defining lattice nodes as the corners of the CPM lattice sites, each lattice node $i$ covered by a CPM cell pulls on every other node $j$ within the same cell, with a force $\vec{F}$ of magnitude proportional to the distance between the nodes,  $\vec{d}_{i,j}$, and $\vec{d}_{i,j}=0$ if line piece $(i,j)$ intersects with the cell boundary (see Ref.~\cite{Lemmon:2010ju}, Figure S1A-D and Supporting Material for details). The total force $\vec{F}_i$ on node $i$ then becomes,
\begin{equation}\label{eq:LemmonRomer}
\vec{F}_i=\mu\sum_j\vec{d}_{i,j}, 
\end{equation}
In accordance with the assumption that the cytoskeleton has uniform contractility, the line pieces have a constant tension per unit distance $\mu$ \cite{Lemmon:2010ju}. For convex cells, the resultant forces point towards the cell's center of mass. For non-convex cells, the resultant forces are directed towards the individual, convex compartments that the cell shape is composed of (see Supplementary Material). 

\section*{Results}
This work proposes a computational model for the collective response of cells to uniaxial stretch in compliant tissues. In the model, cells apply contractile forces onto a compliant substrate. The resulting strains in the matrix affect the motility of the cell itself and the motility of its neighbors. In all of the simulations described in this paper, we stretched a substrate of Young's modulus 12kPa with a stress of $\sigma_\mathrm{stretch}=1000$ N/m$^2$ applied to the boundary of the matrix in the FEM. This results in a static strain of around 8\% on the matrix. The cellular tension $\mu$ (see Eq.~\ref{eq:LemmonRomer}) was set to 0.0025 nN/$\mu$m, resulting in local strains around the tips of elongated cells of up to 2\%, amplifying the static strain to values of around 10\%. 

\subsection*{Individual cell response to uniaxial stretch is amplified by cell contractility}
\begin{figure}[H]
\centering
\includegraphics{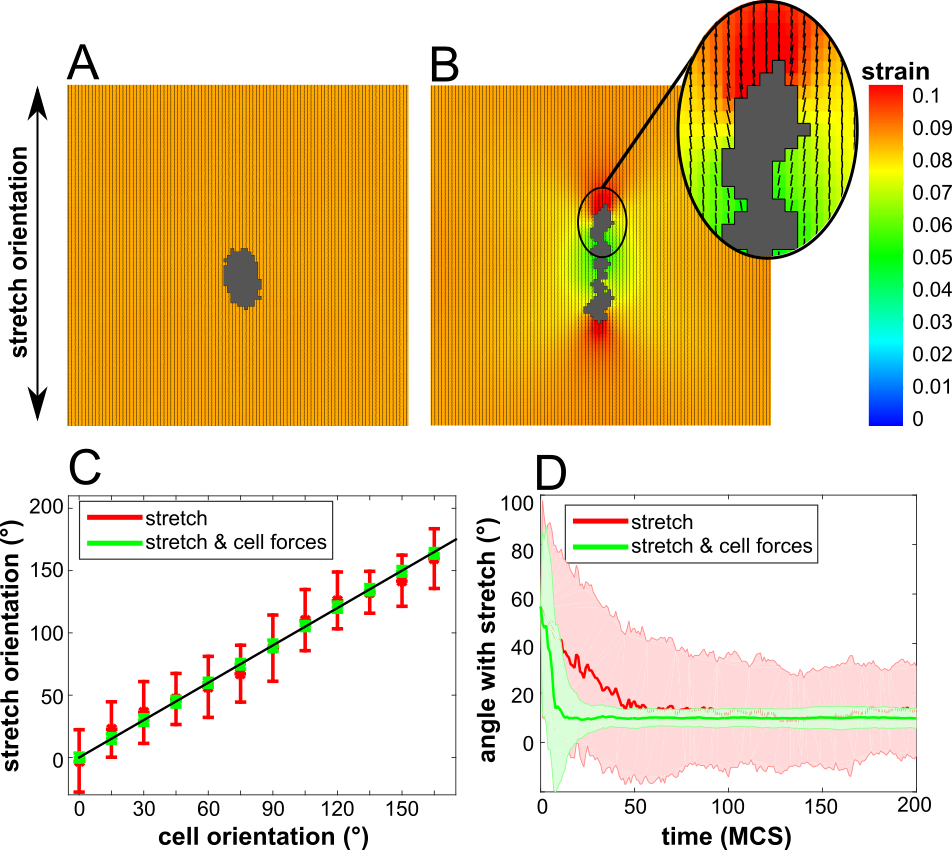}
\caption{(A) Non-contractile cell on substrate stretched along 0$^\circ$ at 500 MCS; (B) contractile cell on substrate stretched along 0$^\circ$ at 500 MCS; (A-B) colors: principal strain magnitude; orientation and length of black line pieces: orientation and magnitude of principal strain; (C) cell orientation as a function of stretch orientation at 500 MCS; averaged over $n=100$ simulations; error bars: standard deviations; black line shows linear fit; (D) time series of the orientation of a single cell on a substrate stretched over 0$^\circ$, averaged over $n=100$ simulations; shaded regions: standard deviations. Color coding (C-D):  red (with large standard deviations): non-contractile cells; green (with larger standard deviations): contractile cells.}
\label{fig:fig2}
\end{figure}

In order to elucidate how cell traction forces affect individual cell response to uniaxial stretch in our model, we simulated the response of a single cell to uniaxial stretch applied in the vertical orientation. This was carried out both in the presence ($\mu>0$) and in the absence ($\mu=0$) of active cell contraction. Figure~\ref{fig:fig2}A shows a non-contractile cell on a uniaxially stretched ECM after 500 MCS; the cell elongates slightly along the stretch orientation, in accordance with our previous results \cite{Oers:2014pcb}. Figure~\ref{fig:fig2}B shows the same simulation set-up in the presence of active cell contraction. The contractile cell elongates more strongly than the non-contractile cell (Figure~\ref{fig:fig2}A). Interestingly, the cell orients itself along the strain orientation, despite the fact that the contractile forces (Eq.~\ref{eq:LemmonRomer}) counteract the uniaxial stretch. The current choice for $\Delta x$ is based on balance between precision and computation time. To confirm that the model is scalable, we repeated the simulation on grids that were refined by a factor of two ($\Delta x = 1.25 \mu\mathrm{m}$, Figure S3A) and four ($\Delta x = 0.625 \mu\mathrm{m}$, Figure S3B), and observed qualitatively similar behavior.  Out of the batch of 100 simulations, in 38 of the simulations, the middle part of contractile cells became rather slender (Figure S2C), resulting in a cell shape that seems unrealistic, as elongated cells are typically reported to have a spindle-like shape. The area conservation (Eq.~\ref{eq:H}) imposes that extensions are, on average, balanced by retractions. Because contractile cells reduce the uniaxial stretch around the center of the cell, retractions are most likely to occur here, resulting in a slender middle part.

To study single cell orientation in more detail and check the isotropy of the model, we performed 100 simulations of single contractile and non-contractile cells for 500 MCS, using stretch angles in the range $0^\circ$ to $180^\circ$ with increments of $15^\circ$ on a 100 $\times$ 100 lattice, representing a piece of tissue of 250 $\times$ 250 $\mu$m. Cells with a diameter of seven lattice sites were initiated in the middle of the matrix. 
The cell orientation was estimated from the inertia tensor of the cells  (see Supporting Material). Figure~\ref{fig:fig2}C plots the cell orientation as a function of the orientation of stretch for cells without active contraction (red boxes) and with active contraction (green boxes). In both conditions, the cells follow the strain orientation on average. However, the cells that apply active contractile forces on the matrix followed the orientation with much higher accuracy, as evidenced by the much smaller standard deviations. Also, the eccentricities of the contractile cells were much more narrowly distributed than those of non-contractile cells (Figure S2A). Figure~\ref{fig:fig2}D shows that the contractile cells oriented more quickly to the stretch orientation than the non-contractile cells. This behavior was only observed on matrices of intermediate stiffness (Figure S2B). On soft substrates, cells remain small \cite{Oers:2014pcb} and as a result do not apply sufficient force on the matrix. On a very stiff matrix, cells protrude in all directions \cite{Oers:2014pcb} and thus they cannot orient along a specific angle. 

Altogether, the simulated contractile cells aligned with the strain more accurately than the non-contractile cells. This can be explained by a positive feedback loop between cell shape, cell traction forces and strain stiffening. Cells elongate slightly in response to uniaxial stretch. Due to the anisotropic cell shape, cells pull harder on the matrix around the tip of the cells, since the distance between the tip of the cell and the cell interior increases (see Eq.~\ref{eq:LemmonRomer}). So, the matrix stiffens around the tip of the cell which further promotes cell elongation along uniaxial stretch.

\subsection*{Cell contractility enables cells to align with each other in parallel to uniaxial stretch}
\begin{figure}[H]
\centering
\includegraphics{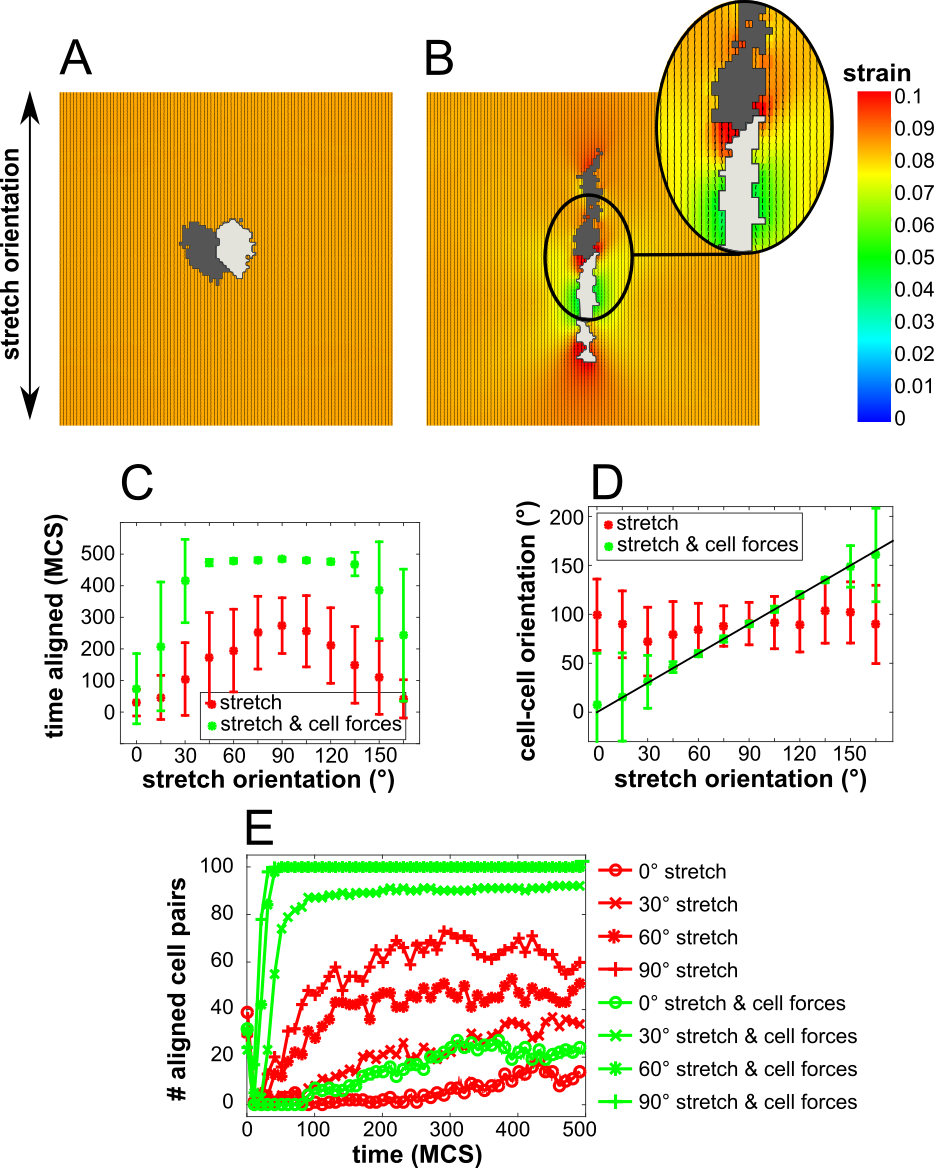}
\caption{(A) Non-contractile cell pair on substrate stretched along 0$^\circ$ at 500 MCS; (B) contractile cell pair on substrate stretched along 0$^\circ$ at 500 MCS; (A-B) colors: principal strain magnitude; orientation and length of black line pieces: orientation and magnitude of principal strain; (C) fraction of time a cell pair is aligned, averaged over $n=100$ simulations, upper bars for contractile cells and lower bars for non-contractile cells; (D) angle of the line connecting the center of masses as a function of stretch orientation at 500 MCS, averaged over $n=100$ simulations, contractile cells follow the extra plotted linear line piece. (C-D): error bars: standard deviations; (E) time series of the number of cell pairs that are aligned on stretched substrates with different stretch orientations; symbols are circle: 0$^\circ$, cross: 30$^\circ$, star: 60$^\circ$, plus: 90$^\circ$, upper lines for contractile cells and lower lines for non-contractile cells. Color coding (C-E):  red: non-contractile cells; green: contractile cells.}
\label{fig:fig3}	
\end{figure}

We next looked at the alignment of neighboring  cells in uniaxially stretched matrices. We simulated the response of two circular cells placed horizontally next to each other on a substrate with a static strain along the vertical axis, both in the presence ($\mu>0$) and in the absence ($\mu=0$) of active cell contraction. Figure~\ref{fig:fig3}A shows a pair of cells on a statically stretched matrix at 500 MCS; the cells elongate slightly and do not align in a head-to-tail fashion. Figure~\ref{fig:fig3}B shows the same simulation set-up in the presence of active cell contraction. In contrast to non-contractile cells, a pair of contractile cells assume a head-to-tail configuration. Also, similar to the response of a single cell found in the previous section, both cells elongate more strongly than the non-contractile cell in Figure~\ref{fig:fig3}A. Notably, the pair of contractile cells assume a head-to-tail configuration along the orientation of uniaxial stretch.

To study this head-to-tail alignment in more detail, we performed 100 simulations of paired cells for 500 MCS for both scenarios on a 200 $\times$ 200 lattice, corresponding to 500 $\times$ 500 $\mu$m, for stretch angles in the range $0^\circ$ to $180^\circ$ with increments of $15^\circ$. Two cells with a diameter of seven lattice sites were initiated in the middle of the matrix, eight lattice sites apart.
Cell-cell alignment was quantified by evaluating the triangle (A,B,C), where A and B are the center of masses of the two cells and C is the point where the lines describing the orientations of the two cells intersect. Table S2 describes how this triangle is used to decide whether a pair of cells is aligned or not. Figure~\ref{fig:fig3}C plots the fraction of time that cells are aligned on a stretched substrate as a function of stretch orientation for cells without active contraction (red boxes) and with active contraction (green boxes). Contractile cells align more often with each other than non-contractile cells. 
To confirm that cells align along the stretch orientation, we measured the orientation of the line connecting the center of masses of the two cells. Figure~\ref{fig:fig3}D plots this cell-cell angle as a function of stretch orientation; a pair of contractile cells align along stretch, compared to non-contractile cells that stick around their initial position (placed horizontally next to each other) and thus keep their initial alignment angle of $90^\circ$. 

Figure~\ref{fig:fig3}E plots the number of cell pairs that aligned  as a function of time in $n=100$ simulations. This shows that with stretch around 0$^\circ$, cells cannot always align. This is because, after initial elongation, the tips of the cells are not in each other's vicinity, such that the cells cannot sense each other's strain. Interestingly, this phenomenon may provide an explanation for an experimental observation reported by Winer {\em et al.}\cite{Winer:2009jg}. Studying the behavior of endothelial cells on compliant matrices, they observed that cell pairs aligned more on 2\,$\mathrm{mg/ml}$ polyacrylamide gels than on a softer, 1 mg/ml gel on which cell assumed an extremely elongated shape. They hypothesized that this ``extremely elongated shape of the cells and thus the shape of the resulting strain field reduced the probability that a second cell would come in contact with the affected gel''. To test this hypothesis in our model, we increased the probability of a cell to come into contact with the strain field of the other cell, by increasing the cellular temperature $T$. Increasing $T$ increases the probability that a cell makes a protrusion. Figure S4A shows the fraction of time a pair of contractile cells are aligned as a function of $T$ and Figure S4B shows how the number of cell pairs that are aligned depend on $T$. This illustrates that pairs of cells more readily align at higher values of $T$. These simulation results thus match the hypothesis of Winer {\em et al.} \cite{Winer:2009jg}. At motilities larger than approximately $T=20$ cell motility became randomized to the extent that the cells could no longer align.

In summary, in our model pairs of contractile cells aligned in head-to-tail configurations along the orientation of stretch, whereas non-contractile cells oriented with the stretch, but not in a head-to-tail fashion. The bipolar strain fields around the contractile cells were instrumental for this cell-cell alignment.

\subsection*{Cell contractility facilitates the self-organization of cells into strings oriented parallel to uniaxial stretch}
\begin{figure}[H]
\centering
\includegraphics{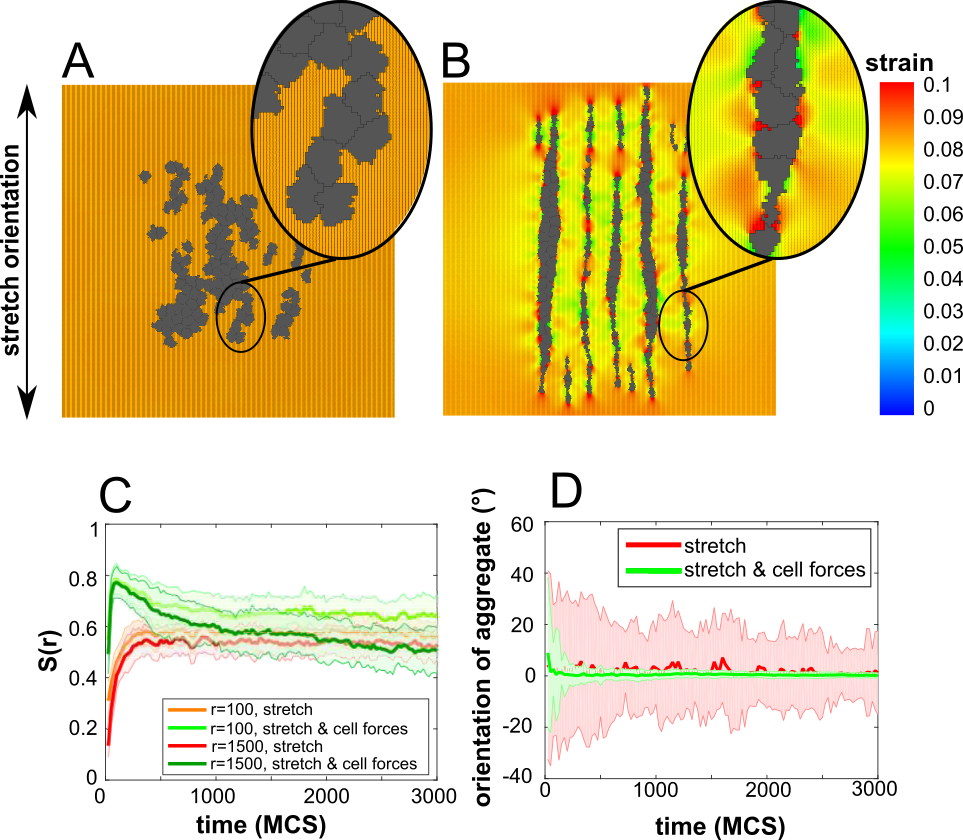}
        \caption{(A) Non-contractile cells on substrate stretched along 0$^\circ$ at 3000 MCS; (B) contractile cells on substrate stretched along 0$^\circ$ at 3000 MCS; (A-B) colors: principal strain magnitude; orientation and length of black line pieces: orientation and magnitude of principal strain;  (C) time series of orientational order parameter $S(r)$, averaged over $n=25$ simulations; color coding: red (lowest line): r=100 $\mu$ m for non-contractile cells, orange (second lowest line): r=1500 $\mu m$ for non-contractile cells, green (highest line): r=100 $\mu m$  for contractile cells, dark-green (second highest line): r=1500 $\mu m$ for contractile cells; (D) time series of the orientation of cell aggregates on a substrate stretched over 0$^\circ$ at 3000 MCS, averaged over $n=25$ simulations; shaded regions: standard deviations. red (with large standard deviations): non-contractile cells; green (with larger standard deviations): contractile cells.}
				\label{fig:fig4}
\end{figure}

After identifying the orientation response of a pair of cells, we asked how cell contractility affects the alignment of a large group of cells. We simulated a group of cells on a stretched matrix, both in the presence ($\mu>0$) and in the absence ($\mu=0$) of active cell contraction. The behavior of the model does not depend on the stretch orientation, so we only show the results for stretching in the vertical orientation in the next sections. Figure~\ref{fig:fig4}A shows a group of cells on a statically stretched matrix in the vertical orientation at 3000 MCS; the cells have elongated slightly and have not migrated away from their initial position.  Figure~\ref{fig:fig4}B shows the same simulation set-up in the presence of active cell contraction. The contractile cell aligned locally with one another in a head-to-tail configuration, as observed in our simulation of paired cells. This cell-cell alignment enables cells to form strings along the orientation of uniaxial stretch, as observed experimentally by Eastwood {\em et al.} \cite{Eastwood:1998cmc}.

To study this behavior in more detail, we performed 25 simulations of a group of cells on a 400 $\times$ 400 lattice, representing a piece of tissue of 1 $\times$ 1 mm, for 3000 MCS for both scenarios, for a stretch angle of $0^\circ$. Cells are initially placed uniformly inside a region of 200 $\times$ 200 lattice sites in the middle of the matrix, as to minimize boundary effects. Cells are initially one lattice site in size. The density of cells was $d=$0.15, yielding around 120 cells. 
To characterize the collective orientation of cells, we measured a two-dimensional orientational order parameter $S(r)$, with range $r$ ($\mu m$), defined for the Cellular Potts Model as in Ref.\cite{Palm2013}. Briefly, $S(r)= \left\langle\cos 2\theta (\vec{X}(s), r)  \right\rangle_s$, where $\vec{X}(s)$ is the center of mass of cell $s$ and $\theta(\vec{X}(s),r)$ is the angle between the orientation of the cell of spin $s$ and a local director, {\em i.e.}, the average orientation of the cells within a radius $r$ around the centroid of cell $s$ (see Supporting Material for detail). $S(r)$ ranges from $S(r)=0$ for configurations of randomly oriented cells, to $S(r)=1$ for fully aligned cells. Figure~\ref{fig:fig4}C plots the orientational order parameter as a function of time, showing a local orientational order ($r=100 \mu m$) for non-contractile cells (orange curve) and for contractile cells (green curve) and the global orientational order ($r=1500 \mu m$) for non-contractile cells (red curve) and for contractile cells (dark-green curve). Contractile cells achieve a higher local and similar global ordering than non-contractile cells. Note that contractile cells initially obtain a high orientational order, close to 0.8. Since cells initially have enough space, they elongate well. When cells start to  adhere to another and form strings, cells in the interior of a string cannot orient well, such that the global orientational order parameter decreases. This is a model artifact which we investigated further in the Supporting Material and address in the discussion section.

To confirm that by contracting the matrix, cells co-align into strings oriented along uniaxial stretch, we measured the orientation of cell aggregates, with a cell aggregate defined as a connected patch of cells (see Supporting Material for details on the calculation). Figure~\ref{fig:fig4}D plots the orientation of cell aggregates as a function of time, of non-contractile cells (green curve) and of contractile cells (red curve). In both conditions, cells form aggregates with an orientation around $0^\circ$, which is the orientation of stretch. The aggregates formed by contractile cells follow the stretch orientation more accurately, as shown by the smaller standard deviations, indicating that strings have formed. 

In our model, contractility facilitates the formation of strings of cells along the stretch orientation, in agreement with experimental observations \cite{Eastwood:1998cmc}. We have shown previously that in unstrained matrices, contractile cells organize into network-like structures \cite{Oers:2014pcb}. We next studied what level of uniaxial stretching is needed for cells to prefer a string-like organization instead of a network-like organization.

\begin{figure}[H]
\centering
\includegraphics{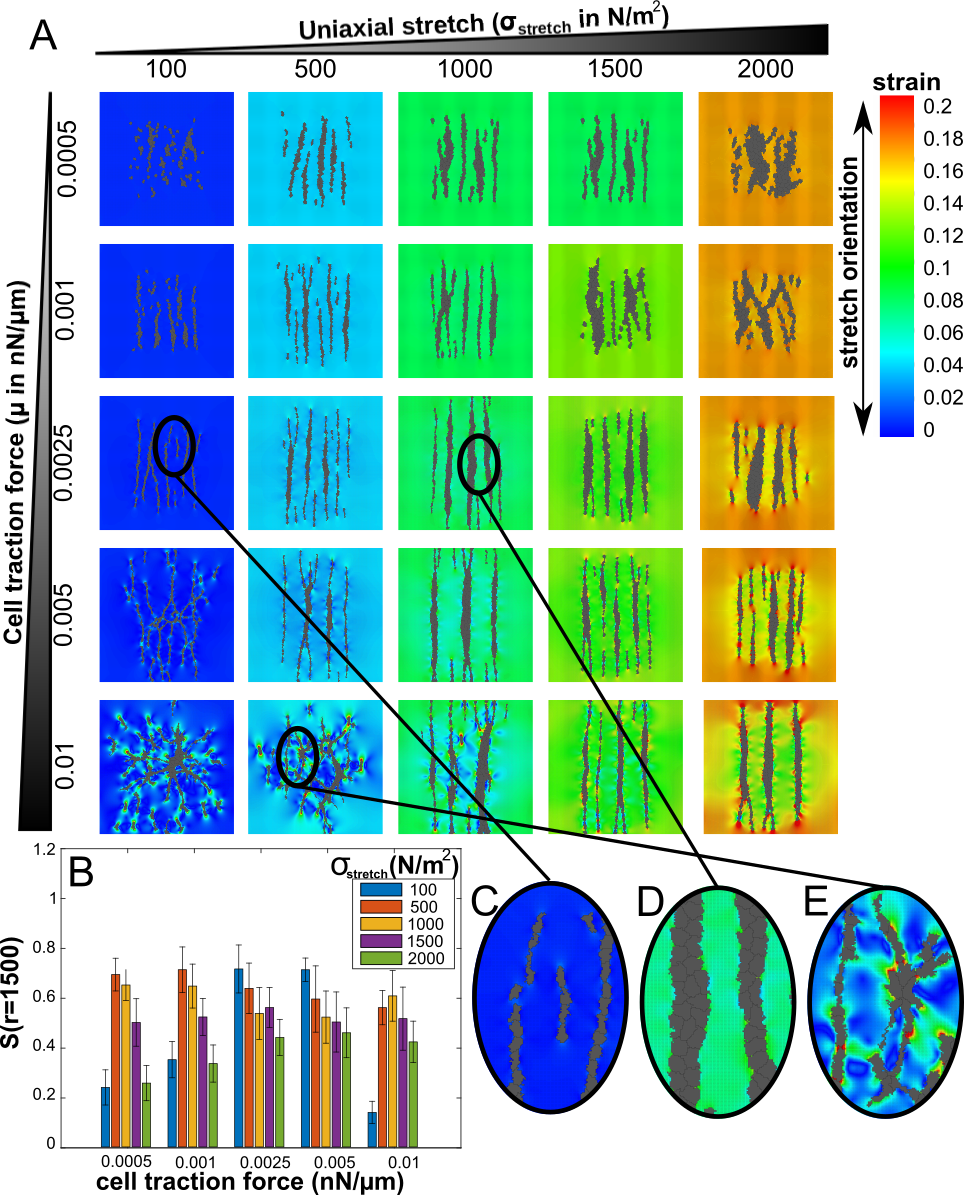}
\caption{(A) Contractile cells on substrate stretched along 0$^\circ$ at 3000 MCS simulated with various values of cell traction force and matrix stretching force; 
(B) Global orientational order parameter $S(r=1500 \mu m)$ at 3000 MCS, averaged over $n=25$ simulations; error bars: standard deviations; bars in bar chart are sorted according to value of $\sigma_\mathrm{stretch}$ as indicated in figure inset;(C) zoom in of cell configuration of $\mu=0.0025$, $F_\mathrm{stretch}=100$; (D) zoom in of cell configuration of $\mu=0.0025$, $F_\mathrm{stretch}=1000$; (E) zoom in of cell configuration of $\mu=0.01$, $F_\mathrm{stretch}=500$. Colors A,(C-E): principal strain magnitude; orientation and length of black line pieces: orientation and magnitude of principal strain. } 
\label{fig:fig5}
\end{figure}

The results of varying uniaxial stretch are shown in Figure~\ref{fig:fig5}. How the amount of uniaxial stretch affects string formation, depends on the magnitude of the cell traction forces. When we varied the uniaxial stretch and fixed the cell traction force to the default parameter value (middle row in Figure~\ref{fig:fig5}A), we observed that cells can more easily align in a head-to-tail configuration and form strings (Figure~\ref{fig:fig5}C) if stretching is lower than the default value (Figure~\ref{fig:fig5}D). Indeed, the global ordering decreases as a function of uniaxial stretch (middle set of barplots in Figure~\ref{fig:fig5}B). In our model, this is explained as follows. Due to the assumed strain stiffening behavior, the cells spread more \cite{Oers:2014pcb} on highly stretched matrices. Then, within strings, cells have less space and orient less well. If cells apply little traction (first row in Figure~\ref{fig:fig5}A), they do not form strings with small uniaxial stretch, but do when stretching is increased. Then, with even more uniaxial stretch, cells orient along stretch, but do not forms strings, similar to non-contractile cells (Figure~\ref{fig:fig4}B). Indeed, the global orientational order parameter shows a  biphasic dependence on uniaxial stretching (first set of bar plots in Figure~\ref{fig:fig5}B). This is explained in our model as follows. Cell forces cannot sufficiently amplify a small uniaxial stretch and thus more uniaxial stretch is needed to instigate string formation. However, at higher uniaxial stretch, the cell traction forces are insufficiently strong to amplify the uniaxial strain and as a result cells do not form strings. Note that these cells do not form networks with little uniaxial stretch, as they do not sufficiently contract the matrix to align with other cells. If cells are highly contractile (last row in Figure~\ref{fig:fig5}A) they form networks, similar to cells on non-stretched matrices (Figure~\ref{fig:modeloverview}F). Higher uniaxial stretching transforms a network into an oriented network (Figure~\ref{fig:fig5}E) and subsequently into strings. Indeed, the global order has a biphasic dependence on stretching (last set of barplots in Figure~\ref{fig:fig5}B). This is because with too little uniaxial stretch, cell generated strains dominate the global strain cue and thus cells do not collectively orient. Of course, if we would increase uniaxial stretch even more, cells would align but not form strings anymore.

To better understand the results in Figure~\ref{fig:fig5}, recall that cells extend towards areas that are stiffened by strain, as described by the sigmoid function $h(E(\epsilon))= 1/(1+\exp (-\beta(E(\epsilon)-E_\theta)))$ (Figure S5B), where  $E(\epsilon)=E_0 ( 1 + (\epsilon /  \epsilon_{st} )$.  Figure S5A shows that cells can only form strings when the matrix is stiffened to values above $E_\theta$ (Figure S5D). If the matrix is not stiffened, or becomes too rigid, the cells will not align (Figure S5C and E). To relate this to Figure~\ref{fig:fig5}, instead of strain, we plotted normalized stiffness values $\frac{E(\epsilon)}{E_\theta}$ in Figure S6. This shows that when the uniaxial stretch stiffens the matrix to values around $E_\theta$ and cell traction forces then stiffen the matrix more, strings can be formed. However, strings can not be formed when the matrix is stiffened too much by either the cells or the uniaxial stretch. 

Altogether, the results suggest that an optimal balance between uniaxial stretch and cell contractility is needed for cells to form strings.  
  
\subsection*{Decreasing cell-cell adhesion promotes string formation in populations with high cell density}
Experimental work has reported two alternative cellular responses to uniaxial strain. Fibroblasts seeded at a density of $10^6$ cells/ml form strings along the orientation of uniaxial stretch \cite{Eastwood:1998cmc}. 3D cultures of endothelial cells at much higher density of $10^7$ cells/ml to $10^9$ cells/ml align along the stretch orientation, but do not form strings \cite{VanderSchaft2011}. Thus, the differences between these two experiments could be due to cellular densities, or due to specific differences between fibroblasts and endothelial cells. In particular, endothelial cells have stronger cell-cell adhesion than fibroblasts, as the endothelial-specific VE-cadherins have stronger bond strengths than the N-cadherins found in fibroblast cell-cell junctions \cite{Panorchan2006}.
\begin{figure}[H]
\centering
\includegraphics{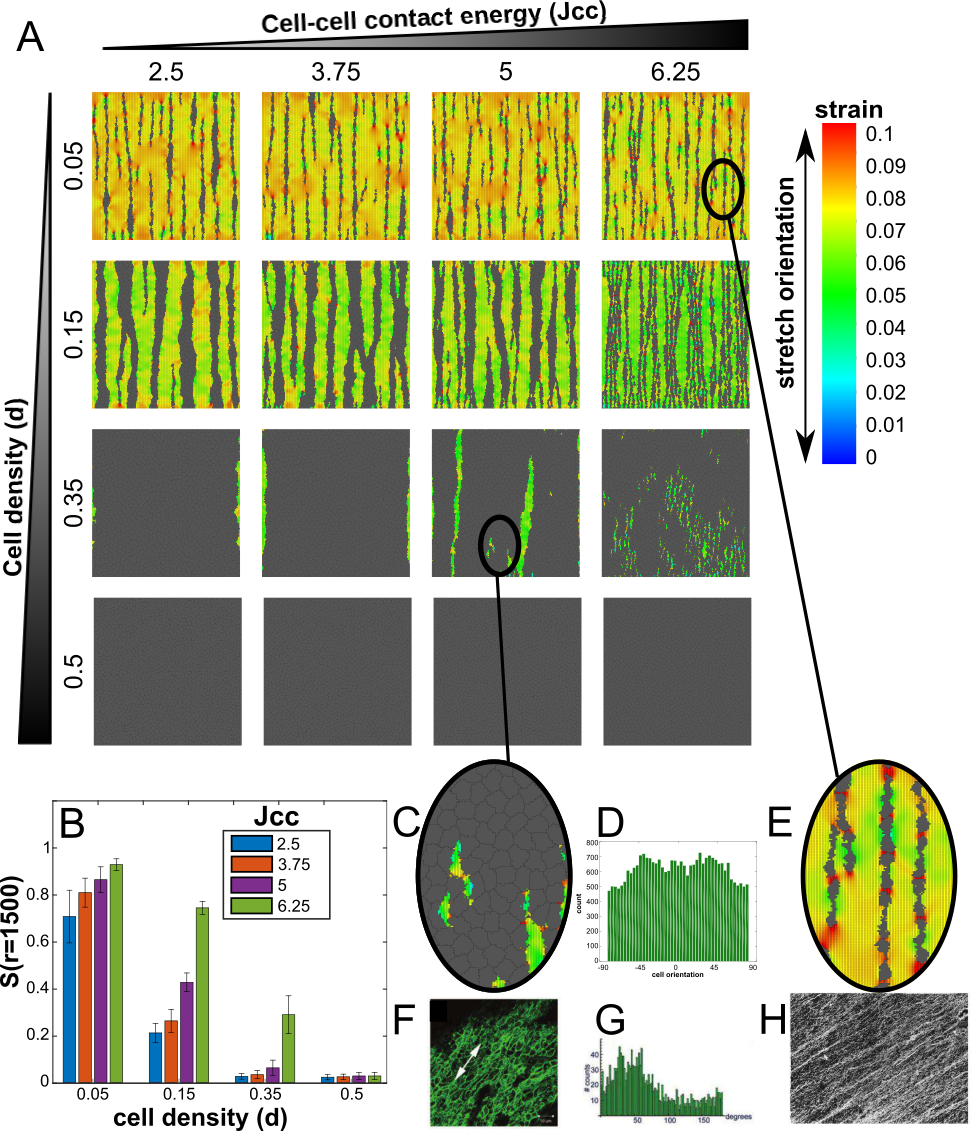}
\caption{(A) Contractile cells on substrate stretched along 0$^\circ$ at 3000 MCS simulated with various values of cell density and cell-cell contact energy; (B) Global orientational order parameter $S(r=1500 \mu m)$ at 3000 MCS, averaged over $n=25$ simulations; error bars: standard deviations, first bar is for the lowest value of $J_\mathrm{cc}$; (B) Global orientational order parameter $S(r=1500 \mu m)$ at 3000 MCS, averaged over $n=25$ simulations; error bars: standard deviations; bars in bar chart are sorted according to value of $J_\mathrm{cc}$ as indicated in figure inset; (C) zoom in of cell configuration of $J_\mathrm{cc}=5$, $d=0.35$; (D) Cell orientations of $J_\mathrm{cc}=6.25$, $d=0.35$; (E) zoom in of cell configuration of $J_\mathrm{cc}=6.25$, $d=0.05$; (F) 3D culture of endothelial cells on uniaxially stretched matrix, taken from Van der Schaft {\em et al.} \cite{VanderSchaft2011}; (G) Orientation of 3D culture of endothelial cells on uniaxially stretched matrix, taken from Van der Schaft {\em et al.} \cite{VanderSchaft2011}; (H) Fibroblasts on uniaxially stretched matrix, taken from Eastwood {\em et al.} \cite{Eastwood:1998cmc}. Colors A,C,E: principal strain magnitude; orientation and length of black line pieces: orientation and magnitude of principal strain. } 
\label{fig:fig6}
\end{figure}
Figure~\ref{fig:fig6}A shows an overview of the final configurations of simulations in which we systematically varied cell density and cell-cell contact energies; Figure~\ref{fig:fig6}B shows the corresponding global orientational order parameters. To better mimic variable densities of cells, we initialized cells on the whole grid of $400 \times 400$ lattice sites. The configurations shown in Figure~\ref{fig:fig6}A suggest that fewer, thicker strings are formed if the cells adhere more strongly to one another (\textit{i.e.}, low $J_{cc}$). Also, the global order parameter increases as the cell-cell contact energies increase (Figure~\ref{fig:fig6}B), suggesting that non-adhering cells respond more easily to the strain cue. At a seeding density $d=0.35$ and mildly repellent cell-cell adhesion settings of $J_{cc}=5$ the final configurations (Figure~\ref{fig:fig6}C) and the distribution of cell orientations (Figure~\ref{fig:fig6}D) qualitatively resemble the experiments by Van der Schaft {\em et al.} \cite{VanderSchaft2011} (Figure~\ref{fig:fig6} F and G). Decreasing cell-cell adhesion and cell densities to $d=0.05$ and $J_{cc}=6.25$ produces configurations similar to Eastwood {\em et al.} \cite{Eastwood:1998cmc} (Figure~\ref{fig:fig6}E and H).  Currently, in completely confluent cell layers with high cell-cell adhesion ($d=0.35,d=0.5$, $J_{cc}=3.75,2.5$ in Figure~\ref{fig:fig6}) the cells do not align at all, because in our model the cells cannot respond to strain at cell-cell interfaces. We investigated this issue further in the Supporting Material and address this in the discussion section.

\section*{Discussion}
In this paper we have presented a computational model to show that active cell contraction can facilitate cellular alignment to the orientation of static uniaxial stretch. The computational model describes motile cells living on top of an elastic substrate, and is based on only a few, experimentally validated assumptions: (a) cells exert contractile forces on the substrate, which locally generate strains in the substrate \cite{Lemmon:2010ju,Califano:2010dp}; (b) cells move by repeatedly attempting to extend or retract pseudopods at random, and (c) along the substrate strain orientation, pseudopod extensions are promoted and pseudopod retractions are inhibited \cite{Pang2011}, a procedure mimicking the maturation of focal adhesions under strain \cite{Pelham1998}. We have shown previously \cite{Oers:2014pcb} that these assumptions suffice to reproduce (1) the elongation of single cells on compliant substrates, (2) the alignment of two adjacent cells, and at the collective level (3) the formation of vascular-like network structures and angiogenesis-like sprouting structures. Here we show that a refined version of this model also reproduces experimentally observed behavior of fibroblasts, endothelial cells and myocytes on statically, uniaxially stretched substrates: (1) cells tend to align in parallel to the uniaxial stretch orientation \cite{Collinsworth:2000cq,Liu2014,VanderSchaft2011} (cf. Figure~\ref{fig:fig2}); (2) cells align with one another in parallel to the uniaxial stretch orientation (Figure~\ref{fig:fig3}); and (3) collectively, the cells form strings oriented along the stretch (Ref.~\cite{Eastwood:1998cmc} and Figure~\ref{fig:fig4}) and they elongate along the stretch in close to confluent layers of cells (Ref.~\cite{VanderSchaft2011} and Figure~\ref{fig:fig6}). Although the assumed response to strains (assumption (c)) makes the simulated cells orient to the stretch without contractility (see Ref.~\cite{Oers:2014pcb} and Figure~\ref{fig:fig2}A), active contractility makes cells elongate more strongly (Figure~\ref{fig:fig2}B), and allows them to respond to strain cues more accurately (Figure~\ref{fig:fig2}C) and more rapidly (Figure~\ref{fig:fig2}D) than non-contractile cells. Thus, a crucial factor for these phenomena is the balance between active cell contractility and the magnitude of the uniaxial stretch cue. Provided the cellular traction forces are sufficiently strong, the cells will collectively organize into oriented strings even in response to very subtle strain cues (Figure~\ref{fig:fig5}). For stronger cell contractility, however, the local strains will override the global strain cue and the cells will organize into network-like patterns as reported previously (see Figure~\ref{fig:fig5}A, lower left panels; also cf.~Ref~\cite{Oers:2014pcb}). The reported model behavior holds for substrates with stiffness of approximately $10\;\mathrm{kPa}$ to approximately $16\;\mathrm{kPa}$, a wider range than the autonomous cell elongation reported previously \cite{Oers:2014pcb}. Note that the exact magnitude of this range depends on the parameter settings and in particular threshold $E_0$ in sigmoid function $h(E(\epsilon))$, whose values were kept unchanged relative to Ref.~\cite{Oers:2014pcb} (see Table S1).  

Experimental validation of our model predictions would need to focus both on the response to uniaxial static stretch of single cells and on the collective behavior of multiple cells. Single cells in our models elongate more easily and reorient more easily to uniaxial static stretch if they contract the matrix. At the multicellular level, contractility induces string formation on uniaxially statically stretched matrices. A number of published \textit{in vitro} experiments already support the single cell behavior that our model predicts. For example, oxidatively modified low density lipoprotein (oxLDL) stimulates the contractility of human aortic endothelial cells, which correlates with increased cell elongation \cite{Byfield2006}. Fibroblasts moving on stretched collagen gels align their trajectories more strongly to the strain orientation than less contractile neutrophils \cite{Haston1983}. To validate single cell response to uniaxial static stretch, we propose experiments in which cells with different contractilities are seeded on a uniaxially stretched matrix as, {\em e.g.}, in Ref.~\cite{Haston1983}. Treatment with lysophosphatidic acid (LPA) can stimulate Rho-mediated contractility \cite{Roy:1999bu,Bischofs2006}, while treating cells with blebbistatin or cytochalasin D inhibits contractility \cite{Winer:2009jg}. At the multicellular level, with increasing uniaxial stretch, our model system switches gradually between networks and strings (Figure~\ref{fig:fig5}). Previous cell culture studies \cite{Eastwood:1998cmc,VanderSchaft2011} have not varied the strain magnitude, but in uniaxially, statically stretched {\em ex ovo} chick chorioallantoic membranes blood vessels realign along stretch \cite{Belle2014}. Further \textit{in vitro} experiments could vary the magnitude of the uniaxial stretch and the degree of contractility using chemical treatments (see {\em e.g.}, Ref.~\cite{Klumpers:2015dy} for a suitable experimental system). The cell density and the cell-cell adhesion strength also influenced the ability of cells to form strings. At high cell densities, simulated cells are less able to form strings, while decreasing cell-cell adhesion restores string formation. Uniaxial stretching experiments where cell-seeding densities are varied and cell-cell adhesion is controlled, by inhibiting or knocking out Cadherins, could validate these predictions.

Although our model is currently not resolved to molecular detail, its simulation results do suggest a mechanistic explanation for the response of cells to static uniaxial stretch. Previous theoretical models \cite{Bischofs2003,De:2007kf} proposed that cells actively regulate their orientation in order to optimize a local mechanical property. Bischofs and Schwarz \cite{Bischofs2003} represented cells as active dipoles, and showed that the dipole can minimize the amount of work required to contract the matrix by orienting along the external strain \cite{Bischofs2003}. This optimization principle was motivated by force-induced focal adhesion maturation: maximum forces will develop at the focal adhesions that are displaced the least. Based on observations suggesting that cells maintain a constant local stress in their microenvironment, De {\em et al.} \cite{De:2007kf} proposed that dipolar cells actively regulate their orientation and contractility in order to maintain a constant optimal amount of local stresses in the matrix. In this model, the dipolar cells reorient to the uniaxial stretch and gradually reduce the magnitude of their contractility in order to reduce the stress in the matrix. Mechanistic rationales certainly motivated these optimization principles, but the mechanisms were not modelled explicitly and a dipole shape was presumed. Our approach instead aims to derive single-cell phenomena and collective cellular responses to strain from a small set of experimentally plausible assumptions at the subcellular level. The present work is only a first step towards this aim. Currently, the local substrate strains regulate the protrusion and retraction probabilities based on a phenomenological function (Eq.~\ref{eq:strain}), which simulates focal adhesion maturation. In our ongoing work we are refining this part of the model by introducing explicit kinetic models of the focal adhesions. 

The current, coarse-grained description has suggested new mechanisms for the experimental observations listed above, but due to a number of technical limitations it still fails to reproduce others. We cannot yet reproduce cell alignment to uniaxial stretch in a completely confluent layer, because the strain-bias of the cell protrusions and retractions is cancelled out at cell-cell interfaces (see Eq.~\ref{eq:strain} and Figures~S7A and~S7B). As a first exploration of the behavior of our model in absence of this effect, we ran a series of simulations in which we differentiated the probability of the retractions relative to extensions. With an decreased retraction probability ($\Delta H_\mathrm{dir}^\mathrm{retraction} = -2 \Delta H_\mathrm{dir}^\mathrm{extension}$), fully confluent cell layers collectively oriented in parallel to stretch (Figures~S7C and~S7D). In contrast, in simulations with an increased retraction probability ($\Delta H_\mathrm{dir}^\mathrm{retraction} = -0.5 \Delta H_\mathrm{dir}^\mathrm{extension}$), the cells oriented themselves perpendicular to the stretch orientations in a confluent layer (Figures~S7E and~S7F). Another result of the absence of strain-effects at cell-cell boundaries, is that contractile cells do not achieve a high global ordering within strings (Figure~\ref{fig:fig4}A); this is because cells in the interior of the strings do not elongate. When the retraction probability is decreased ($\Delta H_\mathrm{dir}^\mathrm{retraction} = -2 \Delta H_\mathrm{dir}^\mathrm{extension}$), the contractile cells reach a higher global ordering ($S(r=1500 \mu m)=0.71$) compared to the non-contractile cells ($S(r=1500 \mu m)=0.51$) (Figure~S8A). In simulations in which the retraction probability is increased ($\Delta H_\mathrm{dir}^\mathrm{retraction} = -0.5 \Delta H_\mathrm{dir}^\mathrm{extension}$), the contractile cells reached a lower global ordering ($S(r=1500 \mu m)=0.37$) compared to the non-contractile cells ($S(r=1500 \mu m)=0.64$), as some cells in the interior of strings started to align perpendicular to strain (Figure~S8B). Despite these quantitative differences, note that cells form strings irrespective of the specific modeling choices (Figure S8 C and D). Also related to this modeling choice is the apparent unrealistic cell shape as presented in Figure S2C. Such cells appear less frequently in simulations where retraction probabilities are decreased ($\Delta H_\mathrm{dir}^\mathrm{retraction} = -2 \Delta H_\mathrm{dir}^\mathrm{extension}$) (Figure S8 C and D)). This work primarily focused on the collective behavior of cells;  in our ongoing work we are developing  more detailed, single-cell models.

Apart from this course-graining of the focal adhesion dynamics and cell motility, our model also relies on other methodological simplifications. The finite-element description of the substrate assumes that the ECM is isotropic, non-fibrous, and linearly elastic. Because of these assumptions, our model best applies to non-fibrous matrices ({\em e.g.}, synthetic polyacrylamide matrices), or to matrices with fibers much smaller than the size of the cells. Of course, more complex matrix mechanics can be modelled using FEM approaches. Interestingly, Aghvami {\em et al.} \cite{Aghvami:2013jbe}, who modelled an anisotropic fiber reinforced material showed similar increased local strains around (non-migratory) cells pulling on uniaxially stretched matrices as in our model. Alternative, agent-based approaches have been proposed for fibrous matrices \cite{Schluter:2012ct,Reinhardt:2013jbe,Reinhardt2014}; in comparison to these models, an advantage of our hybrid approach is in particular its scalability to multicellular systems. As a disadvantage relative to these agent-based approaches, our hybrid set-up relies on an operator splitting approach, which alternates updates of the cell traction forces with the MCS's of the Cellular Potts Model. Although this process speeds up our computations and operator splitting approaches are routinely applied in hybrid modeling (see {\em e.g.}, Refs. \cite{Checa:2014bmm, Albert2014,Albert2016}), it of course also introduces numerical errors: ideally we would recalculate the cellular traction forces and substrate strains after every copy attempt of the CPM. From a biophysical point of view the operator splitting assumption is valid if we can separate the time-scales of the growth and degradation of focal adhesions, such that cell traction forces remain approximately constant during the time represented by one MCS. Indeed, focal adhesion dynamics occur at a timescale of minutes, which is longer than one MCS, which in our model is equivalent to 0.5 to 3 seconds \cite{Oers:2014pcb}. An ongoing improvement of our approach concerns the coupling between the cellular traction forces, as represented by the FMA model (Eq.~\ref{eq:LemmonRomer}), and the representation of these forces in the Hamiltonian (Eqs.~\ref{eq:H}). In the basic CPM, the area conservation and adhesive energy terms in the Hamiltonian describe a pressure and approximate a membrane tension which together represent cell contractility. This allowed us to study the effects of cell-cell contact energies. These terms are not equal to the forces described by the FMA model. The strength of this model is that it produces experimentally validated strain fields. The decoupling of the CPM and the FMA model will become an issue at locations where the two sets of forces are unequal, {\em e.g.}, at cell-cell interfaces  and can affect te mesoscopic cell behavior. Since we are interested in how mesoscopic cell behavior affects the macroscopic level, {\em i.e.}, collective behavior, these approximations and decoupling suffice here. In our ongoing work, we are adopting an approach proposed by Albert and Schwarz \cite{Albert2014} to alleviate this issue.

In summary, we proposed a local cell-matrix feedback mechanism explaining the reorientation of cells to external stretch. In agreement with experimental observations, in this model cell contractility facilitates the reorientation of cells. The proposed mechanism also suffices for the formation of strings along the orientation of stretch. In our future work, we are refining the model by introducing explicit focal adhesion dynamics. This approach will pave the way for issues that our model can currently not explain, including the response of cells to cyclic stretch \cite{Kong2008,Zhong:2011cx}, and the role of cell-substrate adhesivity in the formation of network-like patterns \cite{Califano2008} and collective cell behavior \cite{Albert2016}.

\section*{Supporting citations}

References \cite{Bresenham1965} and \cite{Serra:1983:IAM:1098652} appear in the Supporting Material.

\section*{Author's contributions}

Designed research: EGR, RMHM; performed research and analyzed data: EGR; wrote the manuscript: EGR, RMHM.

\section*{Acknowledgements}

This work is part of the research programme “Innovational Research Incentives Scheme Vidi Cross-divisional 2010 ALW” with project number 864.10.009, which is (partly) financed by the Netherlands Organisation for Scientific Research (NWO). We thank SURFsara (www.surfsara.nl) for the support in using the Lisa Compute Cluster.


\newpage

\begin{section}{Supporting Figures and Tables}

\begin{figure}[H]
\centering
\includegraphics[]{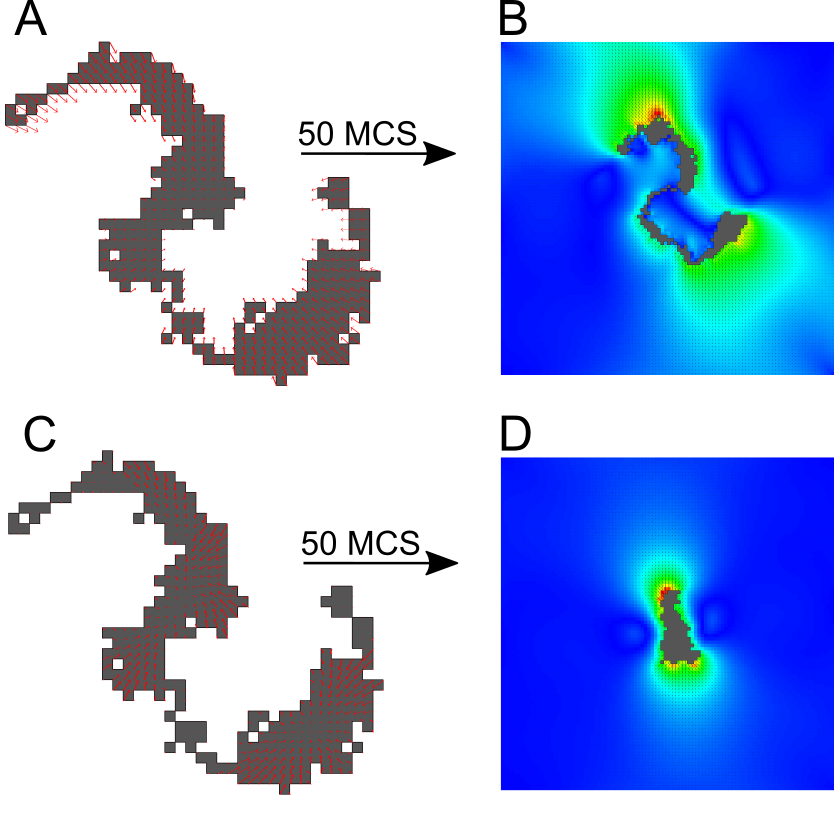}
\caption*{Figure S1: Traction forces for non-convex cell shapes (A) Cell traction forces towards center of mass (van Oers et al. \cite{Oers:2014pcb}); (B) Cell shape 50 MCS after initial configuration in (A)  with $\lambda_{strain}=50$; (C) FMA model \cite{Lemmon:2010ju}; (D) Cell shape 50 MCS after initial configuration in (C) with $\lambda_{strain}=50$; (A-C) length and direction of red arrows: traction force magnitude and direction. B and D colors: principal strain magnitude; orientation and length of black line pieces: orientation and magnitude of principal strain. }

\end{figure}

\begin{figure}[H]
        \centering
        \includegraphics[]{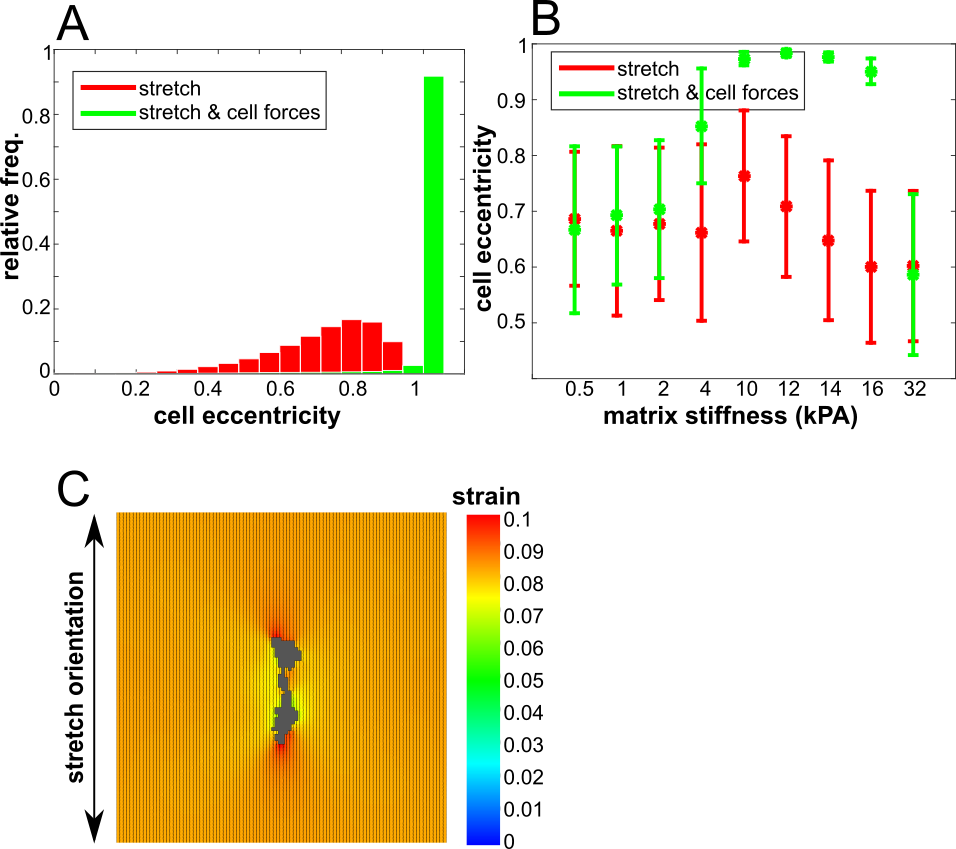}
                \caption*{Figure S2: (A) Distribution of eccentricities of cells on substrate stretched along 0$^\circ$, plotted are the eccentricities of cells during 500 MCS and 100 simulations; (B) Cell orientation as a function of matrix stiffness at 500 MCS; averaged over $n=100$ simulations; error bars: standard deviations. Color coding (A-B): red: non-contractile cells; green: contractile cells; (C) Example of cell shape in which the middle part is much more slender. }
                
\end{figure}

\begin{figure}[H]
\centering
\includegraphics[]{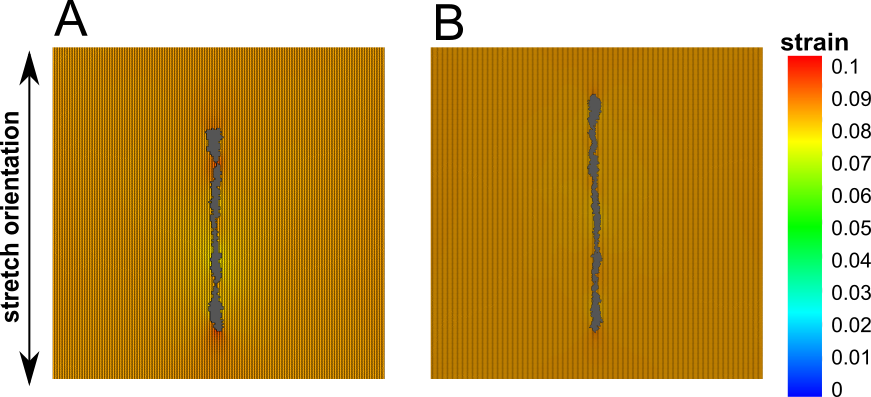}
\caption*{Figure S3: Cells on substrate stretched along 0$^\circ$ at 500 MCS; (A) $\Delta x = 1.25 \mu $m, i.e. refined by a factor of 2 in each lattice direction; (B) $\Delta x = 0.625 \mu $m,i.e. refined by a factor of 4 in each lattice direction.}

\end{figure}

\begin{figure}[H]
\centering
\includegraphics{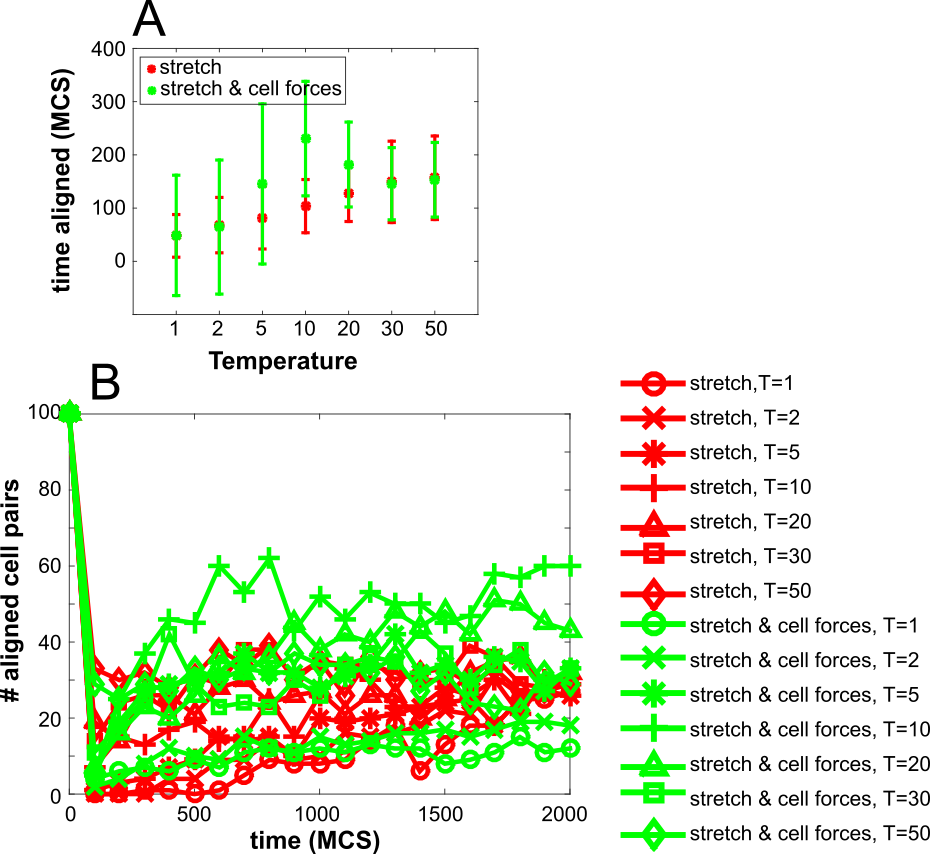}
\caption*{Figure S4: (A) Fraction of time a cell pair is aligned as a function of cellular temperature $T$, averaged over $n=100$ simulations; error bars: standard deviations; (B) time series of the number of cell pairs that are aligned. symbols are circle: $T=1$, cross: $T=2$, star: $T=5$, plus: $T=10$, triangle: $T=20$, square: $T=30$, diamond: $T=50$. Color coding: red: non-contractile cells; green: contractile cells}
\end{figure}

\begin{figure}[H]
\centering
\includegraphics[]{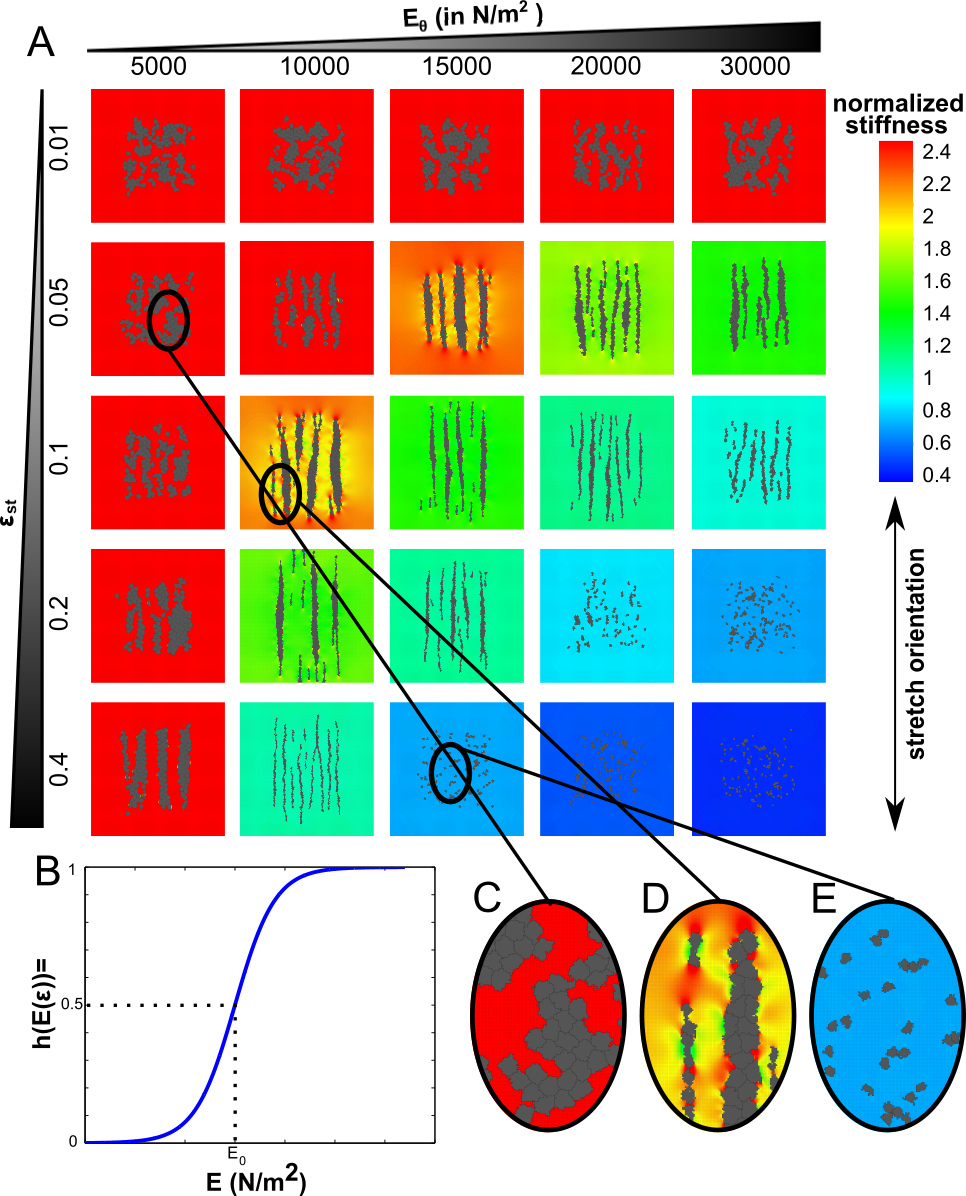}
\caption*{Figure S5: (A) Contractile cells on substrate stretched along 0$^\circ$ at 3000 MCS simulated with various values of $E_\theta$ and $\epsilon_{st}$; (B) graph of $h(E(\epsilon))=1/(1+\exp (-\beta(E(\epsilon)-E_\theta)))$; (C) zoom in of cell configuration of $E_\theta=5000$,$\epsilon_{st}=0.05$; (D) zoom in of cell configuration of $E_\theta=10000$,$\epsilon_{st}=0.1$  (E) zoom in of cell configuration of $E_\theta=15000$,$\epsilon_{st}=0.4$ Colors A,(C-E): normalized stiffness, defined as: $\frac{E(\epsilon)}{E_\theta}= \frac{E_0}{E_\theta}(1+\frac{\epsilon}{\epsilon_{st}})$; orientation and length of black line pieces: orientation and magnitude of principal strain. }

\end{figure}

\begin{figure}[H]
        \centering
        \includegraphics[]{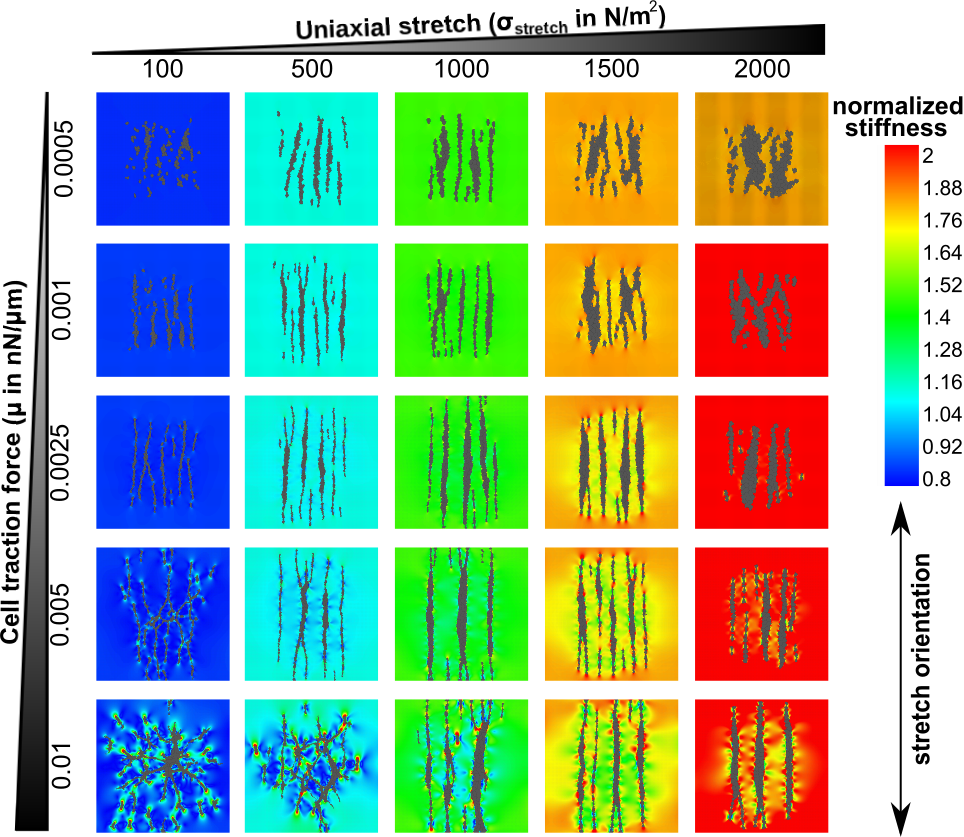}
                \caption*{Figure S6: Contractile cells on substrate stretched along 0$^\circ$ at 3000 MCS simulated with various values of cell traction force and matrix stretching force; Colors: normalized stiffness, defined as: $\frac{E(\epsilon)}{E_\theta}= \frac{E_0}{E_\theta}(1+\frac{\epsilon}{\epsilon_{st}})$; orientation and length of black line pieces: orientation and magnitude of principal strain.}
                
\end{figure}

\begin{figure}[H]
        \centering
        \includegraphics[]{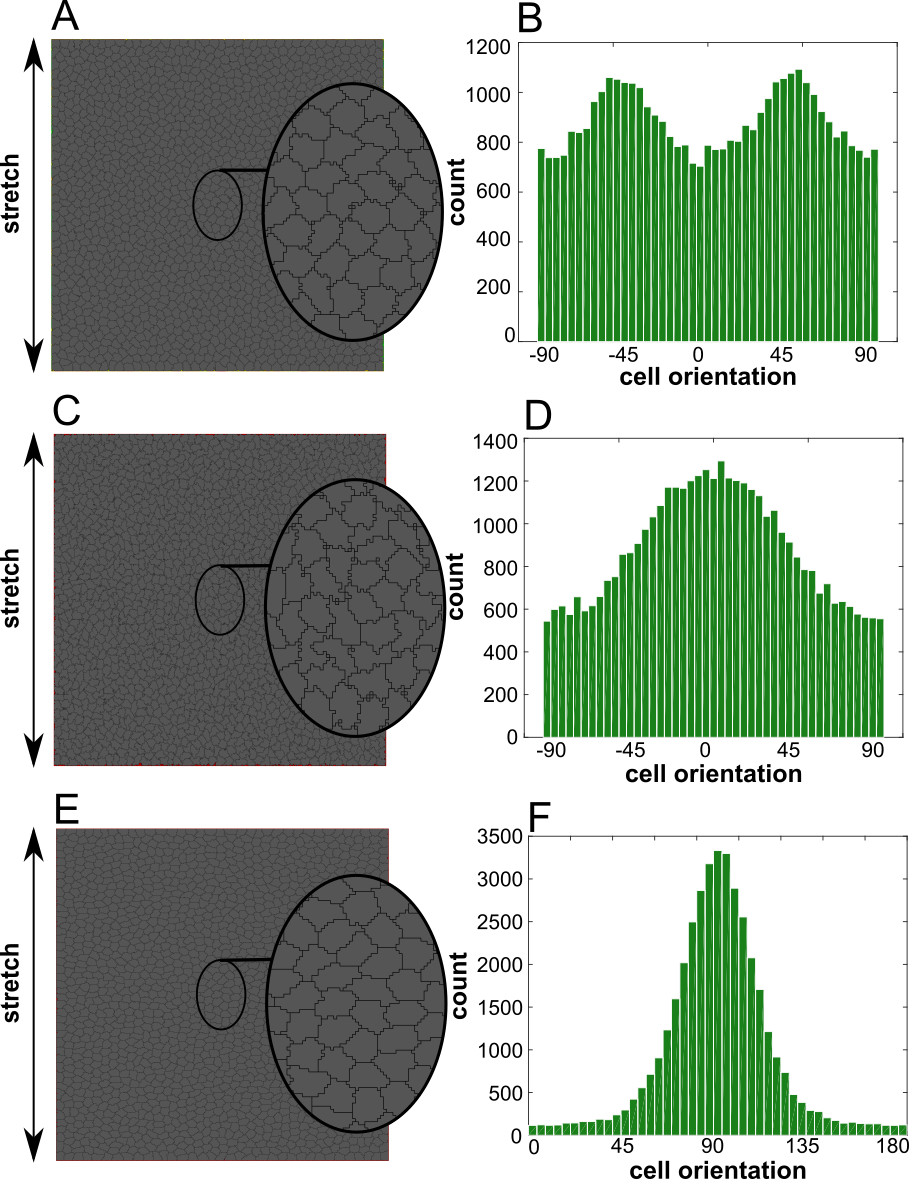}
                \caption*{Figure S7: Contractile cells on substrate stretched along 0$^\circ$ at 3000 MCS with cell density $d=0.5$. (A) Original model; (B) corresponding cell orientations; (C) Model with $\Delta H_\mathrm{strain}^\mathrm{retraction} = -0.5 \Delta H_\mathrm{strain}^\mathrm{extension}$; (D) corresponding cell orientations; (E) Model with $\Delta H_\mathrm{strain}^\mathrm{retraction} = -2 \Delta H_\mathrm{strain}^\mathrm{extension}$; (F) corresponding cell orientations. }
					      
\end{figure}

\begin{figure}[H]
\centering
\includegraphics{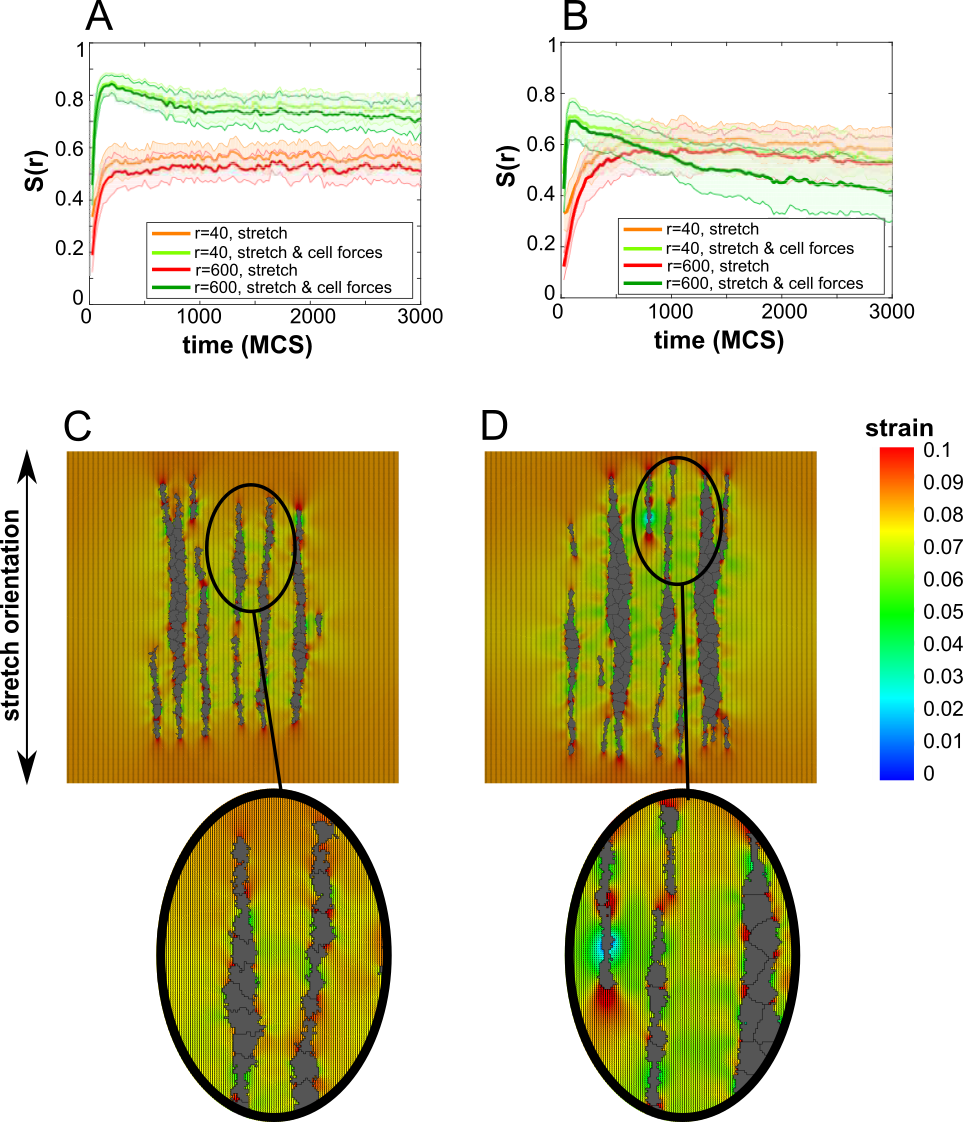}
        \caption*{Figure S8: Time series of orientational order parameter, averaged over $n=25$ simulations; color coding: red: r=40 for non-contractile cells, orange: r=600 for non-contractile cells, green: r=40 for contractile cells, dark-green: r=600 for contractile cells. (A) Model with $\Delta H_\mathrm{strain}^\mathrm{retraction} = -0.5 \Delta H_\mathrm{strain}^\mathrm{extension}$ 
; (B) Model with $\Delta H_\mathrm{strain}^\mathrm{retraction} = -0.5 \Delta H_\mathrm{strain}^\mathrm{extension}$; Contractile cells on substrate stretched along 0$^\circ$ at 3000 MCS. (C) Model with $\Delta H_\mathrm{strain}^\mathrm{retraction} = -0.5 \Delta H_\mathrm{strain}^\mathrm{extension}$ 
; (D) Model with $\Delta H_\mathrm{strain}^\mathrm{retraction} = -0.5 \Delta H_\mathrm{strain}^\mathrm{extension}$.  }
				
\end{figure}

\begin{table}[H]
\centering
\begin{tabular}{|l|l|l|l|}
\hline
\textbf{Parameter symbol} & \textbf{Description} & \textbf{value} & \textbf{units} \\ \hline 
\hline
$ \Delta x $ & width of lattice site & 2.5  & $\mu $m \\ \hline
$A$ & target area & 50 & lattice sites \\ \hline
$J_{cc}$ & cell-cell contact energy & 3.75 & - \\ \hline
$J_{cm}$ & cell-medium contact energy & 1.875 & -\\ \hline
$\lambda$ & strength of volume constraint & 250 & - \\ \hline
$\lambda_{strain}$ & strength of cell response to strain & 24 & - \\ \hline
$T$ & cellular temperature & 5 & -\\ \hline
$\mu$ & cell traction per unit length & 0.0025 & nN $\mu$m$^{-1}$\\ \hline
$E_0$ & Young's modulus & 12 & kPa\\ \hline
$\upsilon$  & Poisson's ratio & 0.45 & -\\ \hline
$\tau$ & substrate thickness & 10 & $\mu$m \\ \hline
$E_\theta$ & threshold for stiffness sensitivity &  15 & kPa\\ \hline
$\beta$ & steepness of stiffness sensitivity & 0.5 & kPa$^{-1}$\\ \hline
$\epsilon_{st}$ & strain stiffening parameter & 0.1 & -\\ \hline
$\sigma_{stretch}$ & uniaxial stretch & 1000 & N/m$^2$\\ \hline
$d$ & cell density & 0.15 & $\frac{\# \lbrace \vec{x}: \sigma(\vec{x})>0 \rbrace}{\# \lbrace \vec{x}: \sigma(\vec{x})=0 \rbrace}$\\ \hline
$r$ & radius for calculating order parameter & 100/1500 & $\mu m$ \\ \hline
\end{tabular}
\caption*{Table S1: Parameters.}
				\label{tab:stab1}

\end{table}

\end{section}

\begin{section}{Alignment quantification details}

\begin{subsection}{Cell elongation and orientation}

To quantify cell elongation and orientation, we used the inertia tensor $I$ of a cell $\sigma$:
\begin{equation}
I(\sigma)=\left(
\begin{matrix}
\sum_{\vec{x}\in C(\sigma)}(x_2-\overline{C}_2(\sigma))^2 & -\sum_{\vec{x}\in C(\sigma)}(x_1-\overline{C}_1(\sigma))(x_2-\overline{C}_2(\sigma)) \\
-\sum_{\vec{x}\in C(\sigma)}(x_1-\overline{C}_1(\sigma))(x_2-\overline{C}_2(\sigma)) & \sum_{\vec{x}\in C(\sigma)}(x_1-\overline{C}_1(\sigma))^2
\end{matrix}\right).
\end{equation}
Here, $\overline{C}(\sigma,t)$, is the center of mass of cell $\sigma$ at MCS (time) $t$, given by
\begin{equation}
\overline{C}(\sigma,t)=\frac{1}{|C(\sigma,t)|}{\sum_{\vec{x}\in C(\sigma,t)}\vec{x}},
\end{equation}
with $C(\sigma,t)$, the set of coordinates of the lattice sites occupied by cell $\sigma$ at MCS $t$.
\\
Cell elongation is quantified by the eccentricity $\xi$ of a cell, given by $\xi(\sigma) = \sqrt{1-\left({\frac{e_1(I(\sigma))}{e_2(I(\sigma))}}\right)^2}$, where $e_1(I(\sigma))\leq e_2(I(\sigma))$ are the eigenvalues of $I(\sigma)$. An eccentricity close to zero corresponds to roughly circular cells and cells with an eccentricity close to unity are more elongated. Further, the orientation of a cell is given by the orientation of the eigenvector associated with the largest eigenvalue of the inertia tensor $I(\sigma)$.

\end{subsection}

\begin{table}[H]
\centering
\begin{tabular}{|l|l|l|l|}
\hline 
 & condition & example & aligned? \\ \hline \hline
case 1 & $\theta_1,\theta_2,\theta_3 < 90 $ &  \raisebox{-\totalheight}{\includegraphics[scale=0.5]{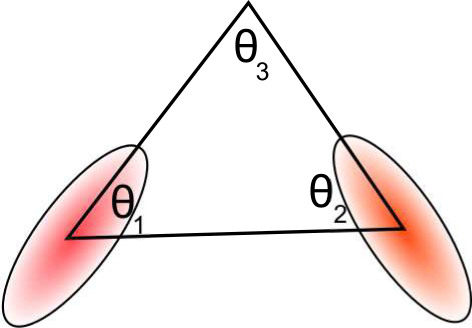}}  & no \\ \hline
case 2 & $\theta_3 \geq 90 $ &\raisebox{-\totalheight}{\includegraphics[scale=0.5]{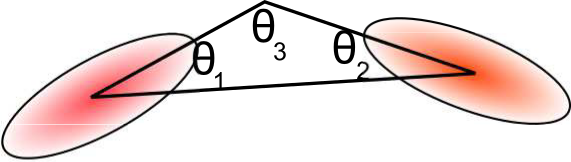}}& yes \\ \hline
 case 3 & $90 < \theta_2 < 135 $ & \raisebox{-\totalheight}{\includegraphics[scale=0.5]{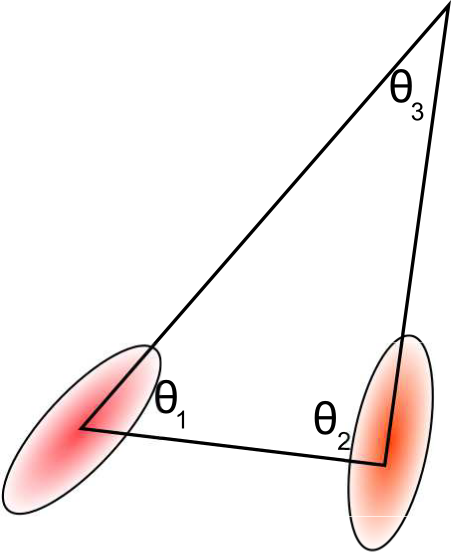}} & no \\ \hline
case 4 & $\theta_2 \geq 135 $ & \raisebox{-\totalheight}{\includegraphics[scale=0.5]{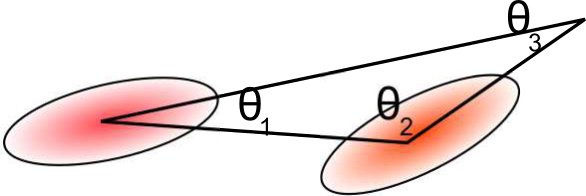}} & yes \\ \hline
\end{tabular}
\caption*{Table S2: Determination of paired cell alignment}
\label{tab:stab2}

\end{table}
			
\begin{subsection}{Orientational order}

For the orientational order parameter $S(r)$ we calculated $\theta(\vec{x},r)$: the angle between the direction of the long axis $\vec{v}(\sigma(\vec{x}))$ of the cell at $\vec{x}$, and a local direction $\vec{n}$, which is the weighted local average of cell orientations, taken within a radius $r$ around $\vec{x}$, such that $\vec{n}(\vec{x},r)=\left < \vec{v}(\sigma(\vec{y})) \right >_{\lbrace \vec{y} \in \mathbb{Z} : |\vec{x}-\vec{y}|<r \rbrace }$. The orientational order parameter is then defined as $S(r)= \left <  \cos 2\theta (\vec{X}(\sigma ), r )  \right > _\sigma $ where $\vec{X}(\sigma )$ is the center of mass of cell $\sigma$.

\end{subsection}

\begin{subsection}{String orientation}

To determine the orientation of strings (or cell aggregates), we first find the connected components of the cell pattern, by applying morphological closing on the pattern \cite{Serra:1983:IAM:1098652}, using a line of five lattice sites with an angle equal to the stretch orientation. We then took the connected components larger than 300 lattice sites and determined the average orientation of those. Aggregate orientations were calculated in the same way as the orientation of a single cell, by using the inertia tensor.

\end{subsection}

\end{section}

\begin{section}{Model details}
\begin{subsection}{Cell traction forces}

In the FMA model \cite{Lemmon:2010ju}, the force $\vec{F}_i$ acting on node $\vec{n}_i$ of a cell is determined by 
\begin{equation}\label{eq:lr}
\vec{F}_i=\mu\sum_{\vec{n}_j} |\vec{n}_i-\vec{n}_j|,
\end{equation}
where the sum is over nodes $\vec{n}_j$ in the same cell of which the straight line connecting node $j$ with node $i$ is completely within the cell. $\mu$ is the cell tension in nN $\mu$m$^{-1}$. To determine whether a line between nodes stays within the cell, one needs to know which lattice sites this line crosses. The Bresenham algorithm \cite{Bresenham1965} is used for this purpose. Now let $s(\vec{n},\vec{n_1})$ be the lattice sites that the line $l(\vec{n},\vec{n_1})$ between node $\vec{n}=(n_x,n_y)$ and node $\vec{n^1}=(n^1_x,n^1_y)$ crosses. In order to be consistent, we impose that $s(-\vec{n_1},\vec{n}) = s(\vec{n},\vec{n_1})$ mirrored vertically and turned 180$^\circ$ clockwise, so that $\vec{n}$ pulls on $\vec{n_1}$ if and only if $\vec{n_1}$ pulls on $\vec{n}$. In order to prevent a bias in either $45^\circ$ or $-45^\circ$, we impose that $s(\vec{n},\vec{n_1})=s(\vec{n},(n^1_x,-n^1_y))$ mirrored horizontally. The resulting lattice sites are shown in an example in Figure S9.
\begin{figure}[H]
	\centering
	\includegraphics[width=0.5\textwidth]{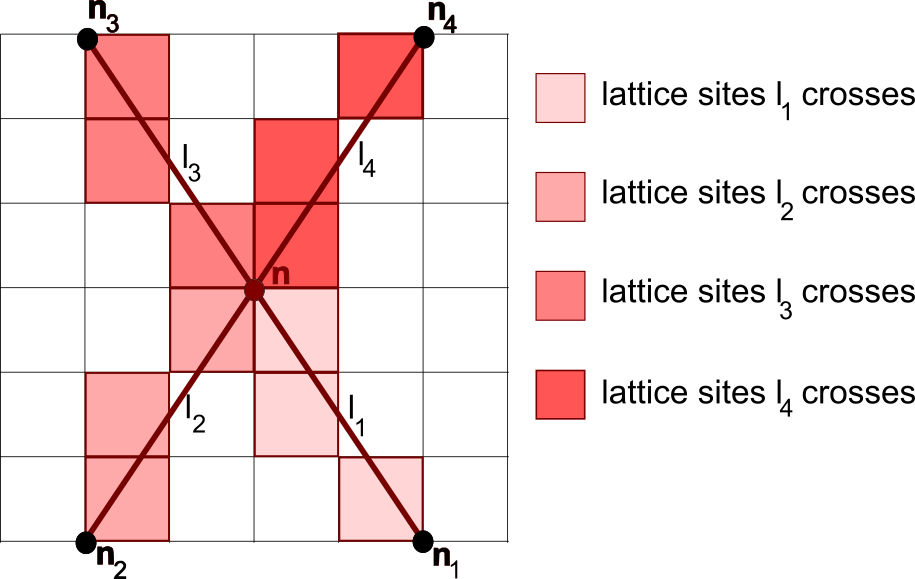}
	\caption*{Figure S9: The lattice sites $s_1$ that the line $l_1$ from node $n$ to $n_1$ crosses is determined using the Bresenham line algorithm. The lattice sites $s_2,s_3,s_4$ corresponding to lines $l_2,l_3,l_4$ from node $n$ to node $n_2,n_3,n_4$ are such that $s_4=s_1$ mirrored horizontally, $s_3=s_1$ mirrored vertically and turned 180$^\circ$ clockwise and $s_2=s_3$ mirrored horizontally. }        
\end{figure}
\end{subsection}

\begin{subsection}{Cellular responses to local strains}
In the calculation of the response of the CPM to the local strains in the substrate, we previously used the strain in the target site when a cell was extending and the strain in the source site was used when a cell was retracting. We changed this assumption to make cell behavior more compatible with focal adhesion dynamics on strained tissues. For an extending cell, i.e. when $\sigma(\vec{x})>0$, the strain in the target site promotes the maturation of a focal adhesion in the protrusion. When a cell is retracting, i.e. $\sigma(\vec{x}\prime)>0$, it costs a lot of energy to unbind a matured focal adhesion from the target site. 

\medskip

In our current model formulation \cite{Oers:2014pcb} cells perceive an increase in  matrix stiffness as a result of compressive ($\epsilon<0$) and extension strains ($\epsilon>0$), while in our original model this was only implemented for extensions strains. This was adapted to avoid a directional bias of cell elongation in $\pm 45^\circ$, see next section.  

\end{subsection}

\begin{subsection}{Single cell orientation in the absence of uniaxial stretch}

A bias in the angle of cell orientations can occur as a result of the square lattice. In Van Oers et al. \cite{Oers:2014pcb}, forces were pointed towards the center of mass (Figure S1A). With this model, a small bias in cell elongation oriented along $\pm 45^\circ$ was found (Figure S10A). We found that the origin of this bias lies in the mechanotaxis term in the Hamiltonian $\Delta H_{\mathrm{strain}}$, that describes a cell perceiving stiffening of the matrix, as a result of positive, stretching strains. When we also let cells perceive strain stiffening as a result of negative strains, i.e. compression, the bias is reduced. This is shown in Figure S10 A and B, in which we plot the orientation of cells on a unstretched matrix with stretch stiffening only and stretch and compression stiffening, respectively. With stretch and compression stiffening, cells are still able to elongate, as a Poisson ratio $\nu < 0.5$ makes sure that stretch strains are higher than compression strains, so cells protrude more preferably towards stretch strains and can thus promote elongation. It is not completely clear to us why the inclusion of compressions stiffening reduces the orientational bias: we found this bias effect by investigating the Hamiltonian for spin copies: with strain stiffening for stretch only, diagonal spin copies gave a higher $dH_{\mathrm{mechanotaxis}}$. 
We discovered that the FMA model (in which nodes only pull on other nodes if they are connected by a straight line within a cell \cite{Lemmon:2010ju}) is another origin for a cell orientation bias, but now along $0^\circ$ and $90^\circ$. We suspect a reason for this, which is illustrated in Figure S11. Our reasoning is as follows. A cell that only experiences contact energy (surface tension) and an area constraint, will obtain a round shape. A cell that elongates wants to stay as round as possible and thus prefers to obtain an ellipse shape. Here we show that an ellipse shaped cell with an orientation of $0^\circ$ has a wider tip than an ellipse orientated along $45^\circ$, because of the 2D grid. A wider tip makes the nodes able to pull on more other nodes, causing more highly strained lattice sites and thus more extensions along $0^\circ$. By increasing the cellular temperature $T$, this bias can sufficiently be reduced. This is shown in Figure S10 C and D, in which we plot the orientation of cells on a unstretched matrix with the FMA model for cell temperature $T=1$ and $T=5$, respectively. Because cells elongate with slightly different parameters for the model with the FMA model compared to previous work (\cite{Oers:2014pcb}), we changed some parameter values with respect to our original work \cite{Oers:2014pcb}. So, in the analysis on cell orientation presented here, we used our previous parameter as in \cite{Oers:2014pcb}:  $J_{cc} = 1.25, J_{cm} = 0.625, \lambda=500,\lambda_{strain} = 20, T=1$.
Finally, there is always a bias in the direction of $\pm 45^\circ$, as a cell elongated in this direction has a lower adhesive energy due to the square lattice. This does not cause major problems as long as sufficient noise is introduced. 

\begin{figure}[H]
        \centering
        \includegraphics[]{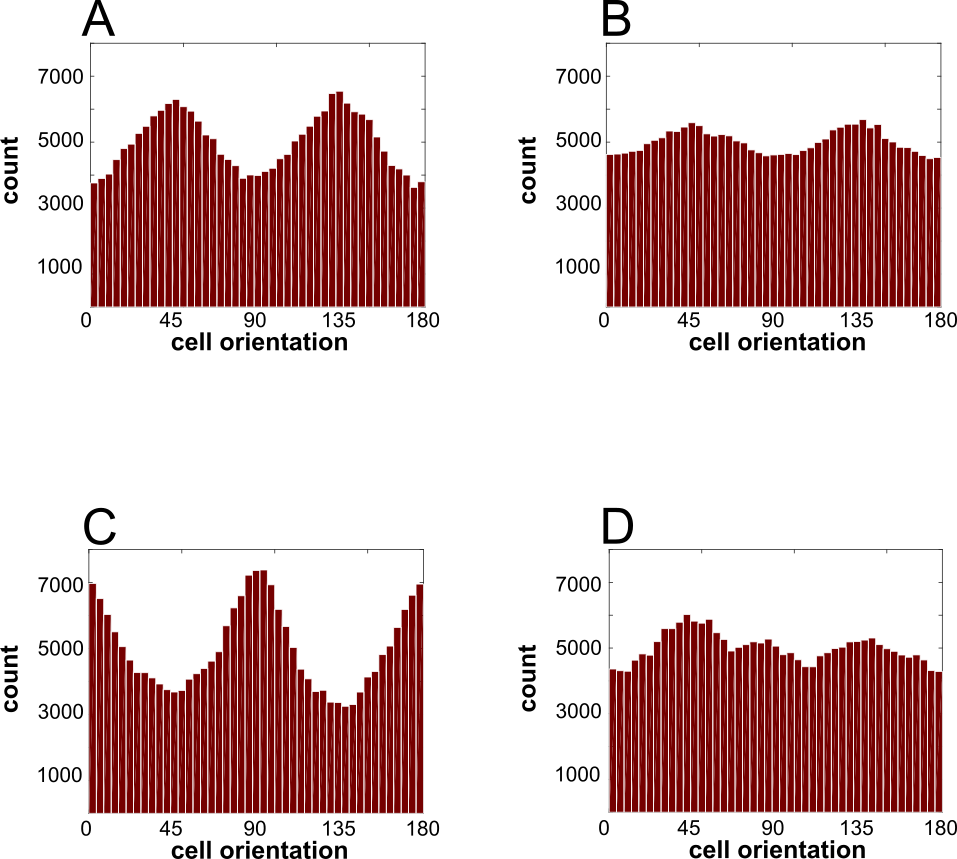}

                \caption*{Figure S10: Cell orientations, one contractile cell on a unstretched substrate, 500 simulations and 500 MCS are plotted. (A) Forces pointed towards center of mass (Van Oers et al. \cite{Oers:2014pcb}), no compression stiffening and $T=1$; (B) Forces pointed towards center of mass (Van Oers et al. \cite{Oers:2014pcb}), compression stiffening and $T=1$; (C) FMA model \cite{Lemmon:2010ju}, compression stiffening and $T=1$; (D) FMA model \cite{Lemmon:2010ju}, compression stiffening and $T=5$.}

\end{figure}

\begin{figure}[H]
        \centering
        \includegraphics[]{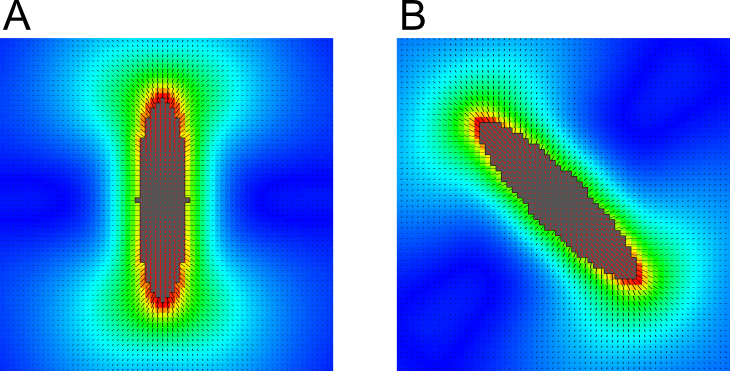}
                \caption*{Figure S11: Strain field around ellipse shaped cell (A) oriented along $0^\circ$; (B) oriented along $45^\circ$.}
\end{figure}
\end{subsection}
\end{section}

\end{document}